\documentclass[aps,prl,twocolumn,groupedaddress,floatfix]{revtex4-1}
\usepackage{amsmath}
\usepackage{amssymb}
\usepackage{graphicx}
\usepackage{float}
\usepackage[english]{babel}
\usepackage{dsfont}
\usepackage{epstopdf}
\usepackage{soul}
\usepackage{color}
\usepackage{braket}
\usepackage[normalem]{ulem}

\usepackage[colorlinks=true,
            linkcolor=magenta,
            urlcolor=blue,
            citecolor=blue]{hyperref}

\definecolor{SMblue}{rgb}{0.5,0.8,1}

\begin{document}
\title{Observation of Topological Band Gap Solitons}
\author{Sebabrata~Mukherjee}
\email{mukherjeesebabrata@gmail.com}
\affiliation{Department of Physics, The Pennsylvania State University, University Park, PA 16802, USA}
\author{Mikael C.~Rechtsman}
\email{mcrworld@gmail.com}
\affiliation{Department of Physics, The Pennsylvania State University, University Park, PA 16802, USA}

\date{\today}

\begin{abstract}
Topological materials exhibit properties dictated by quantised invariants that make them robust against perturbations. This topological protection is a universal wave phenomenon that applies not only in the context of electrons in solid-state materials but also to photonic systems~\cite{raghu2008analogs, wang2009observation, rechtsman2013photonic, hafezi2013imaging}, ultracold atoms~\cite{atala2013direct, jotzu2014experimental}, mechanical systems~\cite{nash2015topological, susstrunk2015observation}, circuits~\cite{ningyuan2015time}, exciton-polaritons~\cite{karzig2015topological, nalitov2015polariton} and beyond. However, the vast majority of research in these systems has focused on the linear domain, i.e., where inter-particle interactions do not play a role. Here, we experimentally observe solitons -- waves that propagate without changing shape as a result of nonlinearity -- in the bulk of a photonic Floquet topological insulator. These solitons exhibit fundamentally different behaviour than solitons in ordinary band gaps in that they execute cyclotron-like orbits that are associated with the topology of the lattice. Specifically, we employ a laser-written waveguide array with periodic variations along the waveguide axis that give rise to non-zero Floquet winding number~\cite{rudner2013anomalous}, where the nonlinearity arises from the optical Kerr effect of the ambient glass. The effect described here is applicable to a range of bosonic systems due to its description by the focusing nonlinear Schr\"odinger equation, i.e., the Gross-Pitaevskii equation with attractive interactions.  
\end{abstract}

\maketitle
Since the discovery of integer quantum Hall effect~\cite{klitzing1980new} and its interpretation in terms of topological properties~\cite{thouless1982quantized}, there has been extensive research into 
exploring exotic topological materials in a wide variety of experimental platforms.
The prediction that quantum Hall-like states could be realised for light propagating in a photonic crystal \cite{raghu2008analogs} has led to wide interest~\cite{lu2014topological,   ozawa2019topological} in the interplay between topological protection and photonic properties, 
especially effects that are not realised  
in the context of electrons in solid-state materials.  After its first observation in a gyromagnetic photonic crystal at microwave frequencies~\cite{wang2009observation}, topological protection has been demonstrated at optical frequencies in synthetic lattices of photonic waveguides~\cite{rechtsman2013photonic} and in the near-infrared range in ring resonators~\cite{hafezi2013imaging}.  The investigation into topological states in electromagnetic systems have been largely limited to the linear domain, where photons propagate independently, governed by Maxwell's equations with linear dielectric and magnetic susceptibilities. In other words, these topological states are described as a system of non-interacting particles with topologically non-trivial energy bands and characterised by integer-valued invariants, such as Chern numbers. 

Among the most fundamental effects in nonlinear optics is the optical Kerr effect: a variation of the material refractive index that is proportional to the local intensity of light. This intensity-dependent refractive index is a manifestation of the nonlinear dielectric polarisation induced by the propagating optical field. In other words, photons with high intensity can effectively interact with one another, mediated by the ambient medium. Indeed, the nonlinear Schr\"odinger equation describing the propagation of light through a nonlinear medium, is equivalent to the Gross-Pitaevskii equation, which describes the interaction between bosons in a Bose-Einstein condensate in the mean-field limit.  Hence, photonic lattices are a natural platform for studying the effects that arise from the interplay of topology and nonlinear dynamics.        

Here, we observe optical spatial solitons \cite{barthelemy1985propagation, christodoulides1988discrete, segev1992spatial, eisenberg1998discrete, stegeman1999optical, fleischer2003observation} in an anomalous photonic Floquet topological insulator \cite{rudner2013anomalous, mukherjee2017experimental, maczewsky2017observation}.  Specifically, we use a lattice of femtosecond-laser-written waveguides that execute a particular modulation of the waveguide paths, which act to effectively break time-reversal 
symmetry in the transverse plane of the lattice.  This lattice has a non-zero Floquet winding number, implying the presence of chiral edge states \cite{rudner2013anomalous}. 
A family of solitons spectrally resides in the topological band gap of the lattice, and during propagation, the peak of the solitons executes cyclotron-like rotations that are inherited from the linear host lattice.
The remainder of the wavefunction rotates around the peak in the same sense.  Like previous theoretical predictions of solitons that reside in topological gaps, it thus shows behaviour that appears to be intrinsic to the topological nature of the system ~\cite{lumer2013self, ablowitz2014linear, leykam2016edge, marzuola2019bulk}.  In that sense, the solitons are of a fundamentally different nature than previously-observed band gap solitons.

\begin{figure*}[t!]
\center
\includegraphics[width=12.5 cm]{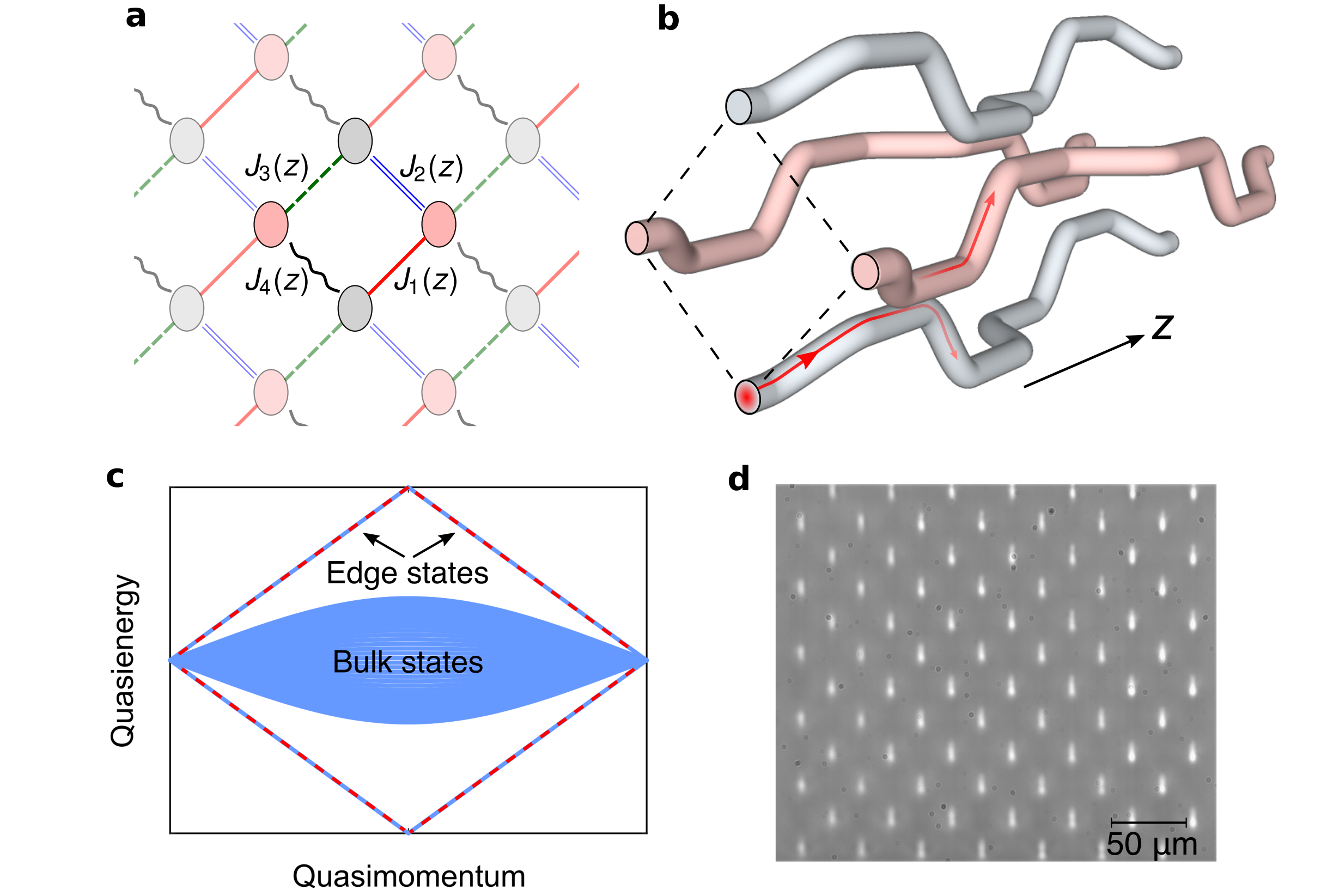}
\caption{{\bf Photonic implementation of an anomalous Floquet topological insulator. a,} Schematic diagram of a periodically-driven square lattice where the couplings $J_m(z)$ [$m\!=\!1,..,4$]
are switched on and off in a cyclic (spatially and $z$-periodic) manner. The propagation distance $z$ plays the role of time in the (nonlinear) Schr\"odinger equation. 
{\bf b,} Simplified sketch showing how this driving protocol is implemented using laser-fabricated 3D waveguide networks. Only four sites are shown here for one complete driving period, $z_0$. {\bf c,} Floquet quasienergy spectrum in the linear regime showing a Chern-zero bulk band and the chiral edge modes. Note that the edge modes connect the top and bottom of the bulk band, which can only happen in the presence of a suitable time-periodic driving. 
{\bf d,} White-light transmission micrograph of the facet of a driven square lattice with 84 waveguides fabricated by femtosecond-laser-writing.
}
\label{fig1}
\end{figure*}

In the presence of the optical Kerr effect, the propagation of light through a photonic lattice with nearest-neighbour evanescent coupling is described by the discrete nonlinear Schr\"odinger equation, under the paraxial approximation:
\begin{eqnarray}
\label{nlse}
i\frac{\partial}{\partial z} \phi_s(z)=\sum_{\left\langle s' \right\rangle} H_{ss'} \phi_{s'} - |\phi_s|^2 \phi_s\; ,
\end{eqnarray}
where the propagation distance ($z$) plays the role of time ($z\! \leftrightarrow \! t$), 
$H_{ss'}$ is the linear tight-binding Hamiltonian (the summation is over neighbouring sites only). We define $|\phi_s|^2\!=\! g|\psi_s|^2$ 
where $|\psi_s|^2$ is the optical power at the $s$-th waveguide and $g$ is a parameter determined by the nonlinear refractive index coefficient $n_2$, the effective area of the waveguide modes and the wavelength of light. At sufficiently low optical power, the nonlinear term of Eq.~\eqref{nlse} can be ignored. Here, we have used the positive or self-focusing nonlinearity (corresponding to attractive interactions in the Gross-Pitaevskii equation), which was experimentally validated for the nonlinear medium studied in this work (see Supplementary Information, Sec.~\hyperref[Char]{3}).
In the absence of optical losses, the total energy and the renormalised power (${\cal P}\equiv\!\sum_s |\phi_s|^2$) are conserved quantities.
Nonlinearity in the off-diagonal coupling term (i.e.,~nonlinearity of evanescent coupling strength) is negligible in our experiments.

Consider a periodically-modulated photonic square lattice~\cite{mukherjee2017experimental, maczewsky2017observation, mukherjee2018state} with nearest-neighbour couplings $J_m(z)$ [$m\!=\!1,..,4$] which are engineered in a cyclic (spatially and $z$-periodic) manner such that every waveguide is coupled to only one of its nearest neighbours at a given propagation distance $z$ (Fig.~\ref{fig1}a-b). The driving period $z_0$ is split into four equal steps and 
over each quarter of the driving period, only one of the four couplings is switched on, with $\int J_m(z) {\text{d}} z \!=\! \Lambda$, where the integral is taken over one coupling operation. 
In the linear regime, the quasienergy spectrum for this periodically-driven lattice can be obtained by diagonalising the Floquet evolution operator over one period, defined as $\hat{U}(z_0)\!=\!\mathcal{T} \exp(-i\int_0^{z_0} \hat{H}(\tilde{z}) {\text{d}}\tilde{z})$, where $\mathcal{T}$ indicates the `time' ordering and $\hat{H}(z)\!=\!\hat{H}(z+z_0)$ is the periodic linear Hamiltonian. 
The width of the bulk band is determined by $\Lambda$: for $\Lambda=\pi/2$, the bulk band becomes perfectly flat. In experiments, we set $\Lambda\!=\!1.85\pm 0.05$, and thus the bulk band is weakly dispersive. 
Fig.~\ref{fig1}c shows the spectrum which was calculated for a strip-geometry aligned along the vertical direction and closed along the horizontal direction, for experimental parameters. 
Due to the periodicity of quasienergy, the edge modes can travel across the energy gap at quasienergy $\varepsilon\!=\!\pi/z_0$ connecting the top and bottom of the band structure.
In other words, a single chiral edge mode exists above and below the bulk band (propagating in the same direction on a given edge) implying that the Chern number of the bulk band is zero.
For such anomalous Floquet topological insulators~\cite{kitagawa2010topological, rudner2013anomalous}, the topology can be captured using a different topological invariant, the winding number.
This scenario can only arise in the presence of a suitable time-periodic driving: anomalous Floquet topological insulators have no analogue in static systems.  Importantly, there is only one band gap in the system (Fig.~\ref{fig1}c), and it is topological.   

\begin{figure*}[t!]
\center
\includegraphics[width=16.0 cm]{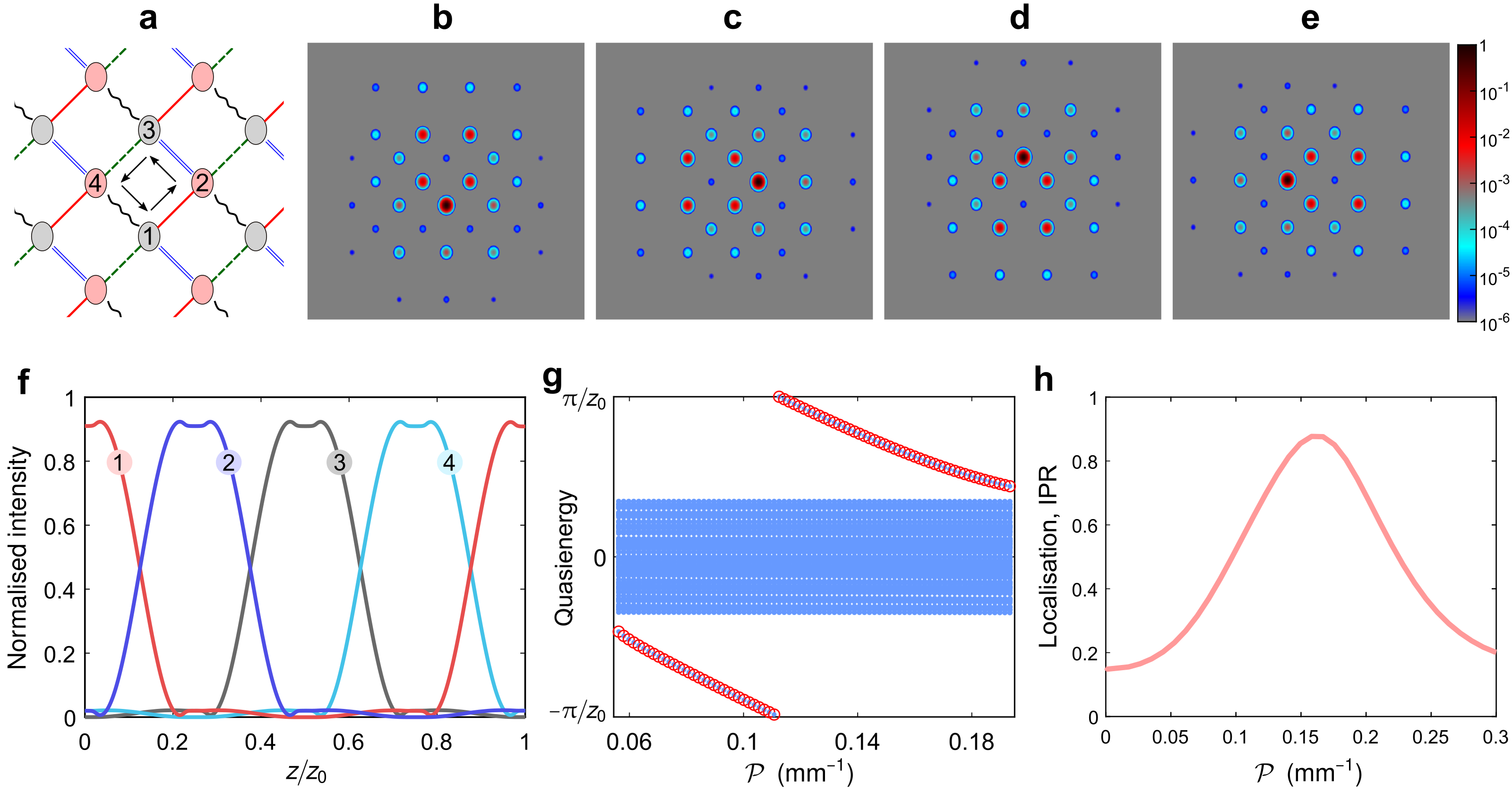}
\caption{{\bf Topological soliton performing cyclotron-like motion. a,} Schematic diagram showing the four sites ($1$-$4$) where the maximum optical power of a topological soliton is contained during propagation.  {\bf b-e,} Normalised intensity profile of a soliton (at ${\cal P}\!=\!0.088$~mm$^{-1}$) rotating counter-clockwise and cycling back to itself after each complete period of driving; the colour map is in log scale. Here, the soliton profile is shown for each quarter of a complete period, i.e.~$z\!=\!\{0, 1, 2, 3\}z_0/4$. {\bf f,} Variation of normalised intensity at the four sites [$1$-$4$ in (a)] showing the dynamics in a complete period. 
{\bf g,} Quasienergy as a function of renormalised power showing the family of bulk solitons (red circles) on both sides of the linear modes. {\bf h,} A signature of these gap solitons can be experimentally detected by launching light into a single bulk waveguide and measuring the degree of localisation of the output intensity patterns. As a function of renormalised power, the output intensity pattern at $z_{\text{}}\!=\!2z_0$ exhibits a distinct localisation feature: a peak in the IPR which is detected in experiments (see Fig.~\ref{fig_Sol2}).
}
\label{Sol_dynamics}
\end{figure*}

Using a self-consistency method modified for Floquet systems~\cite{lumer2013self}, we seek localised nonlinear solutions (solitons) in this modulated photonic square lattice (see Supplementary Information, Sec.~\hyperref[toposol]{5}). The result is self-consistent solitons in the Floquet sense: due to the $z$-periodic driving, the solitons reproduce themselves after each complete period (up to a phase factor), though micro-motion within the Floquet cycle is allowed for. The solitons continuously rotate performing cyclotron-like motion, (see animation1.gif, ref.~\cite{toposolAnimations}).
Figure~\ref{Sol_dynamics}b-e shows the normalised intensity profile (i.e.~$|\phi_s|^2/{\cal P}$) of a soliton at each quarter of a complete period (i.e.~$z\!=\!\{0, 1, 2, 3\}z_0/4$). Figure~\ref{Sol_dynamics}f shows the variation of normalised intensity at the four sites ($1$-$4$ in Fig.~\ref{Sol_dynamics}a) where the maximum optical power of the soliton is contained during propagation.

The quasienergy spectrum is plotted as a function of the renormalised power  
in Fig.~\ref{Sol_dynamics}g, showing a family of gap solitons (red circles) bifurcating from the linear modes (blue). The size (i.e.,~the spatial extent) of the solitons first reduces as a function of power, showing maximal localisation near the mid-gap quasienergy $\pi/z_0$. When the power is further increased, unlike trivial solitons in ordinary static lattices, these Floquet solitons become delocalised (see Supplementary Information, Sec.~\hyperref[toposol]{5}).  
In other words, for a given dispersion of the linear band, the spatial extent of these solitons is determined by their quasienergy -- solitons closer in energy to the linear bulk modes have larger spatial extent (see animation2.gif, ref.~\cite{toposolAnimations}). 
Since the solitons are strongly peaked on a single site, irrespective of their energy, it is possible to probe them in experiments using single-site excitation i.e., by coupling light into a single waveguide of the photonic lattice. Indeed, a signature of these gap solitons can be experimentally detected by measuring the degree of localisation of the output intensity patterns as a function of renormalised power. 
The inverse participation ratio ${\text{IPR}}\! \equiv \!\sum |\phi_s|^4/(\sum |\phi_s|^2)^2$ (a measure of localisation), after propagation of two driving periods, is plotted in Fig.~\ref{Sol_dynamics}h. Here, we observe a clear peak in the IPR, corresponding to the existence of these gap solitons. 
This variation of IPR (i.e.,~delocalisation to localisation to delocalisation) is qualitatively different than in a topologically trivial static lattice, where IPR continuously increases and then saturates at very high nonlinearity (Supplementary Information, Sec.~\hyperref[sol-sqr]{4}).

\begin{figure*}[t!]
\center
\includegraphics[width=16.5 cm]{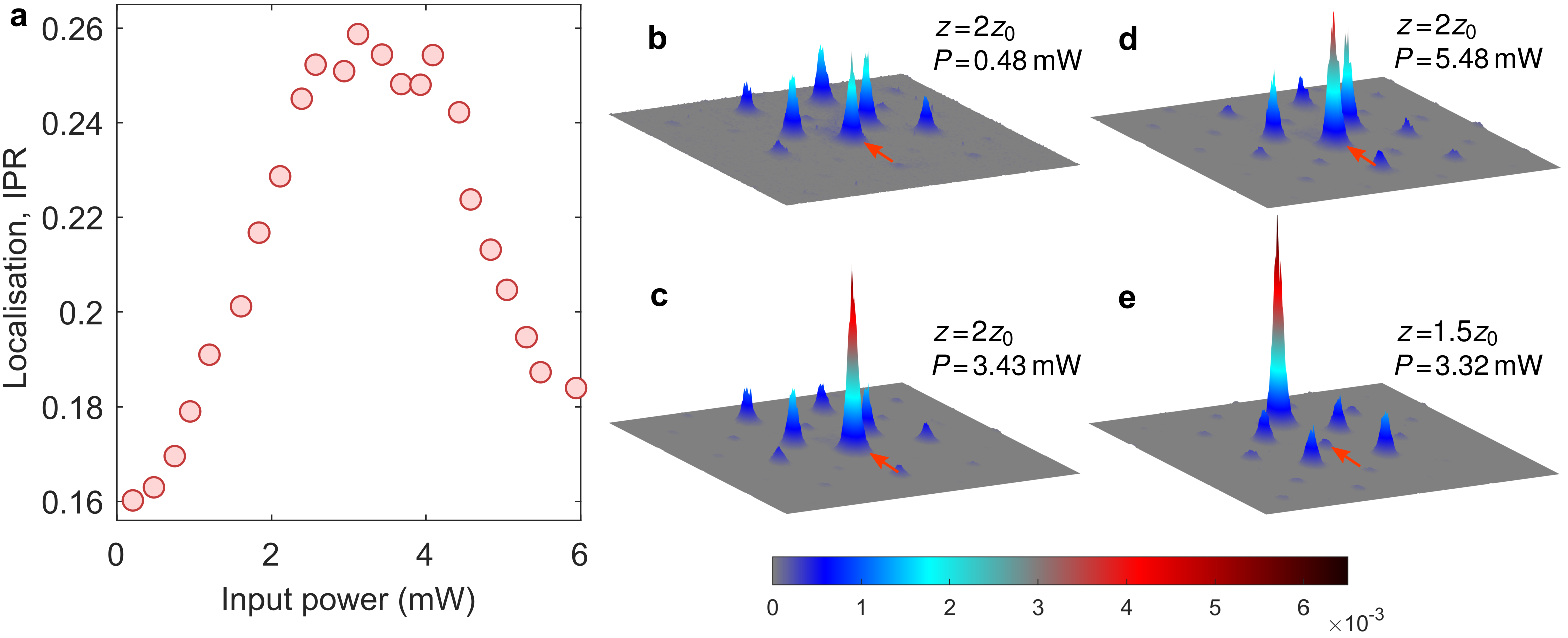}
\caption{{\bf Experimental observation of topological solitons.  a,} Inverse participation ratio as a function of average input power measured at $z\!=\!2z_0$.  The peak corresponds to the existence of the topological band gap solitons.
For these measurements, ${\cal P}$ at the input of the lattice was found to be $0.076$~mm$^{-1}$ per unit average input power in mW.
Corresponding output intensity distributions for three different input powers are shown in {\bf b-d}. The red arrow in each image indicates the site where the light was launched at the input. 
{\bf e,} Most localised output intensity distribution measured at $z\!=\!1.5z_0$. Note that the brightest site in this case is not where the light was launched, corresponding to the cyclotron-like motion of the solitons.
The field of view is smaller than the actual lattice size. Each experimentally-observed intensity pattern is normalised.
}
\label{fig_Sol2}
\end{figure*}

To access a regime for which the Kerr nonlinearity is significant, we coupled intense laser pulses into femtosecond-laser-fabricated waveguide arrays; see Supplementary Information, Sec.~\hyperref[Fab]{1} for fabrication details.
In this situation, $\phi_s$ is a function of both propagation distance and time $t$, $\phi_s\equiv\phi_s(z, t)$. Due to the temporal shape of light pulses, self-phase modulation and chromatic dispersion become relevant (Supplementary Information, Sec.~\hyperref[Char]{2}). 
In our experiments, laser pulses were temporally stretched ({to $t_p\!\approx\!2$~ps) and down-chirped such that the spectral width of the generated wavelengths due to self-phase modulation is $<\!20$~nm after $76$~mm propagation.  
In this wavelength range, the evanescent coupling $J(\lambda)$ only varies by $\Delta J/J\!<\! \pm4$~\%. The variation in the nonlinear parameter $g$ was found to be $\sim\!2$\,\% in this wavelength range. Hence, we can safely ignore the effect of self-phase modulation. 
The effect of chromatic dispersion is negligibly small for the maximal propagation distance ($76$~mm) considered in our experiments since the dispersion length (defined as $t_p^2/|\beta_2|$, $\beta_2$ is the group velocity dispersion) is of the order of hundreds of meters. Additionally, we found that the insertion loss is independent of nonlinearity, implying insignificant nonlinear loss due to multi-photon absorption (Supplementary Information, Sec.~\hyperref[Char]{2}). 

To confirm the validity of these approximations, we first performed an experiment with a topologically trivial static square lattice consisting of straight coupled waveguides. As a function of input power, we observed that the output intensity pattern becomes increasingly localised and finally, all the optical power is trapped largely in the single site where the light was launched at the input (see Supplementary Information, Sec.~\hyperref[sol-sqr]{4} and animation3.gif, ref.~\cite{toposolAnimations}). This baseline experiment clearly demonstrates the formation of highly localised solitons in a topologically trivial energy gap \cite{fleischer2003observation, szameit2006two}.

To experimentally probe the topological gap solitons, a $76$~mm long periodically-modulated square lattice of $84$ sites was fabricated with the previously mentioned driving parameters. A white-light transmission micrograph of the lattice (cross-section) is shown in Fig.~\ref{fig1}d. Each waveguide in the lattice supports the fundamental mode near $1030$~nm wavelength with $8\pm0.2 \; \mu$m and $10.8\pm0.2 \; \mu$m mode field diameters along the horizontal and vertical directions, respectively.  Initially, the waveguides are separated by $26.5 \; \mu$m such that the evanescent couplings are negligibly small. To couple any two desired waveguides, first, the inter-waveguide separation is reduced by synchronously bending the waveguide paths, then the two waveguides are moved straight with $14.5 \; \mu$m fixed centre-to-centre spacing and finally, they are again separated by bending in a reverse manner (Fig.~\ref{fig1}b).

The nonlinear characterisation of this topological photonic lattice is summarised in Fig.~\ref{fig_Sol2}. 
For all measurements, we launched light pulses into a bulk waveguide away from the edges, and during linear diffraction the light does not reach the sides of the array, and thus edge effects can be neglected.  Experimentally, we measured the average optical power, rather than the renormalised power.  
As detailed in the Supplementary Information (Sec.~\hyperref[nl-power]{3}), the renormalised power ${\cal P}$ at the input of the lattice was found to be $0.076$~mm$^{-1}$ per unit average input power in mW.
In the first set of experiments, we measured output intensity distributions at $z\!=\!2z_0$ as a function of average input power (see animation4.gif, ref.~\cite{toposolAnimations}).
The variation of inverse participation ratio with input power is presented in Fig.~\ref{fig_Sol2}a. At low optical power, i.e.,~in the linear regime, this single-site excitation overlaps with the weakly dispersive bulk modes, and light diffracts away from the site in which it is injected (see Fig.~\ref{fig_Sol2}b). 
As input power was increased, output intensity patterns became increasingly localised exhibiting a peak in the IPR near average power $P\!=\!3.4$~mW (see Fig.~\ref{fig_Sol2}c).  Most of the optical power in Fig.~\ref{fig_Sol2}c is contained at the site where the light was initially launched (indicated by the red arrow). When the power was further increased, the output showed a dramatic delocalisation (see Fig.~\ref{fig_Sol2}d), as would be expected from the numerical result presented in~Fig.~\ref{Sol_dynamics}h.

In a second set of experiments, we used a similar photonic lattice with maximum propagation distance  $z_{\text{max}}=1.5z_0$ instead of $2z_0$ (see animation5.gif, ref.~\cite{toposolAnimations}). The inverse participation ratio in this case exhibited a similar behaviour (i.e.,~delocalisation to
localisation to delocalisation) to Fig.~\ref{fig_Sol2}a.
The most localised output intensity pattern was observed near $P\!=\!3.3$~mW (see Fig.~\ref{fig_Sol2}e).
In contrast to Fig.~\ref{fig_Sol2}c, the brightest site in Fig.~\ref{fig_Sol2}e is not located where the light was initially launched, but in another waveguide that is directly across a diagonal from the injected site, which is a direct evidence of the cyclotron-like motion of the solitons. Now comparing Fig.~\ref{Sol_dynamics}h and Fig.~\ref{fig_Sol2}a, the peak of IPR is experimentally observed at a higher power (${\cal P}\!=\!0.26$~mm$^{-1}$) and the contrast of the peak is lower than what is expected from theory. This happens because of linear optical (propagation and bending) losses. Additionally, the front and rear tails (in time) of the pulses always behave linearly, producing a small background and causing a lower contrast of IPR. That said, these effects do not qualitatively change the localisation feature, and thus the observed peak in IPR is a clear experimental signature of the topological band gap solitons.

It is worthwhile to comment on the physical intuition that lies behind our ability to observe the soliton, despite the fact that nonlinear response is relatively weak in the borosilicate-based waveguide array used here.  A key element is our ability to control the `flatness' of the bulk band by tailoring the coupling parameter, $\Lambda$. In solid-state systems, flat bands play a key role in enhancing the relative strength of interactions (a dramatic recent example is twisted bilayer graphene \cite{cao2018unconventional}).  This is also true in the context of photonic waveguide arrays, but in a different way: the width of the linear band sets the power threshold of solitons in two dimensions.  In the extreme case, when the band is perfectly flat, all linear Bloch states are degenerate, and thus localised eigenstates can be constructed as a superposition of these states, implying that stationary states exist even in the linear domain (limit of zero power).  Thus, in this limit, solitons have zero power threshold.  Although the bulk band is not flat for the system described above, the ability to control its width by setting $\Lambda$ sufficiently close to the flat-band limit allows us to observe the soliton at an experimentally accessible value of the power.  

In conclusion, we have experimentally demonstrated a family of Floquet solitons in a topological energy gap. The observation of these topological solitons in our well-controlled experimental setup opens a new avenue towards the investigation of topological nonlinear optics, complementing other platforms such as Rydberg polaritons \cite{clark2019observation} and nonlinear circuits \cite{hadad2016self}. Furthermore, nonlinearity can act as a means to modify \cite{PhysRevLett.121.023901, he_floquet_2019} and probe \cite{kruk2019nonlinear, PhysRevLett.123.103901} topological photonic structures. There are indeed many open questions: what is the nature of the scatter-free of chiral edge states in the presence of nonlinearity? What are the appropriate topological invariants to describe nonlinear systems, and what are their corresponding physical observables? Is the nature of disorder-induced localisation strongly modified by the interplay between interactions and non-trivial topology? We expect that the answers to these questions, among others, will dictate how topological states can be used in useful devices based on wave mechanics, whether in the photonic, acoustic/phononic, optomechanical, atomic, polaritonic or other domains.\\

\noindent{Acknowledgments.$-$} We are pleased to thank Harikumar~K. Chandrasekharan, Jonathan Guglielmon, Daniel Leykam and Jiho Noh for useful discussions and Chris Giebink for use of a supercontinuum laser source for directional coupler characterisation.  SM and MCR gratefully acknowledge support from the Office of Naval Research under award number N00014-18-1-2595, and MCR acknowledges the Packard and Kaufman foundations under fellowship number 2017-66821 and award number KA2017-91788, respectively.

\begin{thebibliography}{44}%
\makeatletter
\providecommand \@ifxundefined [1]{%
 \@ifx{#1\undefined}
}%
\providecommand \@ifnum [1]{%
 \ifnum #1\expandafter \@firstoftwo
 \else \expandafter \@secondoftwo
 \fi
}%
\providecommand \@ifx [1]{%
 \ifx #1\expandafter \@firstoftwo
 \else \expandafter \@secondoftwo
 \fi
}%
\providecommand \natexlab [1]{#1}%
\providecommand \enquote  [1]{``#1''}%
\providecommand \bibnamefont  [1]{#1}%
\providecommand \bibfnamefont [1]{#1}%
\providecommand \citenamefont [1]{#1}%
\providecommand \href@noop [0]{\@secondoftwo}%
\providecommand \href [0]{\begingroup \@sanitize@url \@href}%
\providecommand \@href[1]{\@@startlink{#1}\@@href}%
\providecommand \@@href[1]{\endgroup#1\@@endlink}%
\providecommand \@sanitize@url [0]{\catcode `\\12\catcode `\$12\catcode
  `\&12\catcode `\#12\catcode `\^12\catcode `\_12\catcode `\%12\relax}%
\providecommand \@@startlink[1]{}%
\providecommand \@@endlink[0]{}%
\providecommand \url  [0]{\begingroup\@sanitize@url \@url }%
\providecommand \@url [1]{\endgroup\@href {#1}{\urlprefix }}%
\providecommand \urlprefix  [0]{URL }%
\providecommand \Eprint [0]{\href }%
\providecommand \doibase [0]{http://dx.doi.org/}%
\providecommand \selectlanguage [0]{\@gobble}%
\providecommand \bibinfo  [0]{\@secondoftwo}%
\providecommand \bibfield  [0]{\@secondoftwo}%
\providecommand \translation [1]{[#1]}%
\providecommand \BibitemOpen [0]{}%
\providecommand \bibitemStop [0]{}%
\providecommand \bibitemNoStop [0]{.\EOS\space}%
\providecommand \EOS [0]{\spacefactor3000\relax}%
\providecommand \BibitemShut  [1]{\csname bibitem#1\endcsname}%
\let\auto@bib@innerbib\@empty
\bibitem [{\citenamefont {Raghu}\ and\ \citenamefont
  {Haldane}(2008)}]{raghu2008analogs}%
  \BibitemOpen
  \bibfield  {author} {\bibinfo {author} {\bibfnamefont {S.}~\bibnamefont
  {Raghu}}\ and\ \bibinfo {author} {\bibfnamefont {F.~D.~M.}\ \bibnamefont
  {Haldane}},\ }\href@noop {} {\bibfield  {journal} {\bibinfo  {journal}
  {Physical Review A}\ }\textbf {\bibinfo {volume} {78}},\ \bibinfo {pages}
  {033834} (\bibinfo {year} {2008})}\BibitemShut {NoStop}%
\bibitem [{\citenamefont {Wang}\ \emph {et~al.}(2009)\citenamefont {Wang},
  \citenamefont {Chong}, \citenamefont {Joannopoulos},\ and\ \citenamefont
  {Solja{\v{c}}i{\'c}}}]{wang2009observation}%
  \BibitemOpen
  \bibfield  {author} {\bibinfo {author} {\bibfnamefont {Z.}~\bibnamefont
  {Wang}}, \bibinfo {author} {\bibfnamefont {Y.}~\bibnamefont {Chong}},
  \bibinfo {author} {\bibfnamefont {J.~D.}\ \bibnamefont {Joannopoulos}}, \
  and\ \bibinfo {author} {\bibfnamefont {M.}~\bibnamefont
  {Solja{\v{c}}i{\'c}}},\ }\href@noop {} {\bibfield  {journal} {\bibinfo
  {journal} {Nature}\ }\textbf {\bibinfo {volume} {461}},\ \bibinfo {pages}
  {772} (\bibinfo {year} {2009})}\BibitemShut {NoStop}%
\bibitem [{\citenamefont {Rechtsman}\ \emph {et~al.}(2013)\citenamefont
  {Rechtsman}, \citenamefont {Zeuner}, \citenamefont {Plotnik}, \citenamefont
  {Lumer}, \citenamefont {Podolsky}, \citenamefont {Dreisow}, \citenamefont
  {Nolte}, \citenamefont {Segev},\ and\ \citenamefont
  {Szameit}}]{rechtsman2013photonic}%
  \BibitemOpen
  \bibfield  {author} {\bibinfo {author} {\bibfnamefont {M.~C.}\ \bibnamefont
  {Rechtsman}}, \bibinfo {author} {\bibfnamefont {J.~M.}\ \bibnamefont
  {Zeuner}}, \bibinfo {author} {\bibfnamefont {Y.}~\bibnamefont {Plotnik}},
  \bibinfo {author} {\bibfnamefont {Y.}~\bibnamefont {Lumer}}, \bibinfo
  {author} {\bibfnamefont {D.}~\bibnamefont {Podolsky}}, \bibinfo {author}
  {\bibfnamefont {F.}~\bibnamefont {Dreisow}}, \bibinfo {author} {\bibfnamefont
  {S.}~\bibnamefont {Nolte}}, \bibinfo {author} {\bibfnamefont
  {M.}~\bibnamefont {Segev}}, \ and\ \bibinfo {author} {\bibfnamefont
  {A.}~\bibnamefont {Szameit}},\ }\href@noop {} {\bibfield  {journal} {\bibinfo
   {journal} {Nature}\ }\textbf {\bibinfo {volume} {496}},\ \bibinfo {pages}
  {196} (\bibinfo {year} {2013})}\BibitemShut {NoStop}%
\bibitem [{\citenamefont {Hafezi}\ \emph {et~al.}(2013)\citenamefont {Hafezi},
  \citenamefont {Mittal}, \citenamefont {Fan}, \citenamefont {Migdall},\ and\
  \citenamefont {Taylor}}]{hafezi2013imaging}%
  \BibitemOpen
  \bibfield  {author} {\bibinfo {author} {\bibfnamefont {M.}~\bibnamefont
  {Hafezi}}, \bibinfo {author} {\bibfnamefont {S.}~\bibnamefont {Mittal}},
  \bibinfo {author} {\bibfnamefont {J.}~\bibnamefont {Fan}}, \bibinfo {author}
  {\bibfnamefont {A.}~\bibnamefont {Migdall}}, \ and\ \bibinfo {author}
  {\bibfnamefont {J.}~\bibnamefont {Taylor}},\ }\href@noop {} {\bibfield
  {journal} {\bibinfo  {journal} {Nature Photonics}\ }\textbf {\bibinfo
  {volume} {7}},\ \bibinfo {pages} {1001} (\bibinfo {year} {2013})}\BibitemShut
  {NoStop}%
\bibitem [{\citenamefont {Atala}\ \emph {et~al.}(2013)\citenamefont {Atala},
  \citenamefont {Aidelsburger}, \citenamefont {Barreiro}, \citenamefont
  {Abanin}, \citenamefont {Kitagawa}, \citenamefont {Demler},\ and\
  \citenamefont {Bloch}}]{atala2013direct}%
  \BibitemOpen
  \bibfield  {author} {\bibinfo {author} {\bibfnamefont {M.}~\bibnamefont
  {Atala}}, \bibinfo {author} {\bibfnamefont {M.}~\bibnamefont {Aidelsburger}},
  \bibinfo {author} {\bibfnamefont {J.~T.}\ \bibnamefont {Barreiro}}, \bibinfo
  {author} {\bibfnamefont {D.}~\bibnamefont {Abanin}}, \bibinfo {author}
  {\bibfnamefont {T.}~\bibnamefont {Kitagawa}}, \bibinfo {author}
  {\bibfnamefont {E.}~\bibnamefont {Demler}}, \ and\ \bibinfo {author}
  {\bibfnamefont {I.}~\bibnamefont {Bloch}},\ }\href@noop {} {\bibfield
  {journal} {\bibinfo  {journal} {Nature Physics}\ }\textbf {\bibinfo {volume}
  {9}},\ \bibinfo {pages} {795} (\bibinfo {year} {2013})}\BibitemShut {NoStop}%
\bibitem [{\citenamefont {Jotzu}\ \emph {et~al.}(2014)\citenamefont {Jotzu},
  \citenamefont {Messer}, \citenamefont {Desbuquois}, \citenamefont {Lebrat},
  \citenamefont {Uehlinger}, \citenamefont {Greif},\ and\ \citenamefont
  {Esslinger}}]{jotzu2014experimental}%
  \BibitemOpen
  \bibfield  {author} {\bibinfo {author} {\bibfnamefont {G.}~\bibnamefont
  {Jotzu}}, \bibinfo {author} {\bibfnamefont {M.}~\bibnamefont {Messer}},
  \bibinfo {author} {\bibfnamefont {R.}~\bibnamefont {Desbuquois}}, \bibinfo
  {author} {\bibfnamefont {M.}~\bibnamefont {Lebrat}}, \bibinfo {author}
  {\bibfnamefont {T.}~\bibnamefont {Uehlinger}}, \bibinfo {author}
  {\bibfnamefont {D.}~\bibnamefont {Greif}}, \ and\ \bibinfo {author}
  {\bibfnamefont {T.}~\bibnamefont {Esslinger}},\ }\href@noop {} {\bibfield
  {journal} {\bibinfo  {journal} {Nature}\ }\textbf {\bibinfo {volume} {515}},\
  \bibinfo {pages} {237} (\bibinfo {year} {2014})}\BibitemShut {NoStop}%
\bibitem [{\citenamefont {Nash}\ \emph {et~al.}(2015)\citenamefont {Nash},
  \citenamefont {Kleckner}, \citenamefont {Read}, \citenamefont {Vitelli},
  \citenamefont {Turner},\ and\ \citenamefont {Irvine}}]{nash2015topological}%
  \BibitemOpen
  \bibfield  {author} {\bibinfo {author} {\bibfnamefont {L.~M.}\ \bibnamefont
  {Nash}}, \bibinfo {author} {\bibfnamefont {D.}~\bibnamefont {Kleckner}},
  \bibinfo {author} {\bibfnamefont {A.}~\bibnamefont {Read}}, \bibinfo {author}
  {\bibfnamefont {V.}~\bibnamefont {Vitelli}}, \bibinfo {author} {\bibfnamefont
  {A.~M.}\ \bibnamefont {Turner}}, \ and\ \bibinfo {author} {\bibfnamefont
  {W.~T.}\ \bibnamefont {Irvine}},\ }\href@noop {} {\bibfield  {journal}
  {\bibinfo  {journal} {Proceedings of the National Academy of Sciences}\
  }\textbf {\bibinfo {volume} {112}},\ \bibinfo {pages} {14495} (\bibinfo
  {year} {2015})}\BibitemShut {NoStop}%
\bibitem [{\citenamefont {S{\"u}sstrunk}\ and\ \citenamefont
  {Huber}(2015)}]{susstrunk2015observation}%
  \BibitemOpen
  \bibfield  {author} {\bibinfo {author} {\bibfnamefont {R.}~\bibnamefont
  {S{\"u}sstrunk}}\ and\ \bibinfo {author} {\bibfnamefont {S.~D.}\ \bibnamefont
  {Huber}},\ }\href@noop {} {\bibfield  {journal} {\bibinfo  {journal}
  {Science}\ }\textbf {\bibinfo {volume} {349}},\ \bibinfo {pages} {47}
  (\bibinfo {year} {2015})}\BibitemShut {NoStop}%
\bibitem [{\citenamefont {Ningyuan}\ \emph {et~al.}(2015)\citenamefont
  {Ningyuan}, \citenamefont {Owens}, \citenamefont {Sommer}, \citenamefont
  {Schuster},\ and\ \citenamefont {Simon}}]{ningyuan2015time}%
  \BibitemOpen
  \bibfield  {author} {\bibinfo {author} {\bibfnamefont {J.}~\bibnamefont
  {Ningyuan}}, \bibinfo {author} {\bibfnamefont {C.}~\bibnamefont {Owens}},
  \bibinfo {author} {\bibfnamefont {A.}~\bibnamefont {Sommer}}, \bibinfo
  {author} {\bibfnamefont {D.}~\bibnamefont {Schuster}}, \ and\ \bibinfo
  {author} {\bibfnamefont {J.}~\bibnamefont {Simon}},\ }\href@noop {}
  {\bibfield  {journal} {\bibinfo  {journal} {Physical Review X}\ }\textbf
  {\bibinfo {volume} {5}},\ \bibinfo {pages} {021031} (\bibinfo {year}
  {2015})}\BibitemShut {NoStop}%
\bibitem [{\citenamefont {Karzig}\ \emph {et~al.}(2015)\citenamefont {Karzig},
  \citenamefont {Bardyn}, \citenamefont {Lindner},\ and\ \citenamefont
  {Refael}}]{karzig2015topological}%
  \BibitemOpen
  \bibfield  {author} {\bibinfo {author} {\bibfnamefont {T.}~\bibnamefont
  {Karzig}}, \bibinfo {author} {\bibfnamefont {C.-E.}\ \bibnamefont {Bardyn}},
  \bibinfo {author} {\bibfnamefont {N.~H.}\ \bibnamefont {Lindner}}, \ and\
  \bibinfo {author} {\bibfnamefont {G.}~\bibnamefont {Refael}},\ }\href@noop {}
  {\bibfield  {journal} {\bibinfo  {journal} {Physical Review X}\ }\textbf
  {\bibinfo {volume} {5}},\ \bibinfo {pages} {031001} (\bibinfo {year}
  {2015})}\BibitemShut {NoStop}%
\bibitem [{\citenamefont {Nalitov}\ \emph {et~al.}(2015)\citenamefont
  {Nalitov}, \citenamefont {Solnyshkov},\ and\ \citenamefont
  {Malpuech}}]{nalitov2015polariton}%
  \BibitemOpen
  \bibfield  {author} {\bibinfo {author} {\bibfnamefont {A.}~\bibnamefont
  {Nalitov}}, \bibinfo {author} {\bibfnamefont {D.}~\bibnamefont {Solnyshkov}},
  \ and\ \bibinfo {author} {\bibfnamefont {G.}~\bibnamefont {Malpuech}},\
  }\href@noop {} {\bibfield  {journal} {\bibinfo  {journal} {Physical Review
  Letters}\ }\textbf {\bibinfo {volume} {114}},\ \bibinfo {pages} {116401}
  (\bibinfo {year} {2015})}\BibitemShut {NoStop}%
\bibitem [{\citenamefont {Rudner}\ \emph {et~al.}(2013)\citenamefont {Rudner},
  \citenamefont {Lindner}, \citenamefont {Berg},\ and\ \citenamefont
  {Levin}}]{rudner2013anomalous}%
  \BibitemOpen
  \bibfield  {author} {\bibinfo {author} {\bibfnamefont {M.~S.}\ \bibnamefont
  {Rudner}}, \bibinfo {author} {\bibfnamefont {N.~H.}\ \bibnamefont {Lindner}},
  \bibinfo {author} {\bibfnamefont {E.}~\bibnamefont {Berg}}, \ and\ \bibinfo
  {author} {\bibfnamefont {M.}~\bibnamefont {Levin}},\ }\href@noop {}
  {\bibfield  {journal} {\bibinfo  {journal} {Physical Review X}\ }\textbf
  {\bibinfo {volume} {3}},\ \bibinfo {pages} {031005} (\bibinfo {year}
  {2013})}\BibitemShut {NoStop}%
\bibitem [{\citenamefont {Klitzing}\ \emph {et~al.}(1980)\citenamefont
  {Klitzing}, \citenamefont {Dorda},\ and\ \citenamefont
  {Pepper}}]{klitzing1980new}%
  \BibitemOpen
  \bibfield  {author} {\bibinfo {author} {\bibfnamefont {K.~v.}\ \bibnamefont
  {Klitzing}}, \bibinfo {author} {\bibfnamefont {G.}~\bibnamefont {Dorda}}, \
  and\ \bibinfo {author} {\bibfnamefont {M.}~\bibnamefont {Pepper}},\
  }\href@noop {} {\bibfield  {journal} {\bibinfo  {journal} {Physical Review
  Letters}\ }\textbf {\bibinfo {volume} {45}},\ \bibinfo {pages} {494}
  (\bibinfo {year} {1980})}\BibitemShut {NoStop}%
\bibitem [{\citenamefont {Thouless}\ \emph {et~al.}(1982)\citenamefont
  {Thouless}, \citenamefont {Kohmoto}, \citenamefont {Nightingale},\ and\
  \citenamefont {den Nijs}}]{thouless1982quantized}%
  \BibitemOpen
  \bibfield  {author} {\bibinfo {author} {\bibfnamefont {D.~J.}\ \bibnamefont
  {Thouless}}, \bibinfo {author} {\bibfnamefont {M.}~\bibnamefont {Kohmoto}},
  \bibinfo {author} {\bibfnamefont {M.~P.}\ \bibnamefont {Nightingale}}, \ and\
  \bibinfo {author} {\bibfnamefont {M.}~\bibnamefont {den Nijs}},\ }\href@noop
  {} {\bibfield  {journal} {\bibinfo  {journal} {Physical Review Letters}\
  }\textbf {\bibinfo {volume} {49}},\ \bibinfo {pages} {405} (\bibinfo {year}
  {1982})}\BibitemShut {NoStop}%
\bibitem [{\citenamefont {Lu}\ \emph {et~al.}(2014)\citenamefont {Lu},
  \citenamefont {Joannopoulos},\ and\ \citenamefont
  {Solja{\v{c}}i{\'c}}}]{lu2014topological}%
  \BibitemOpen
  \bibfield  {author} {\bibinfo {author} {\bibfnamefont {L.}~\bibnamefont
  {Lu}}, \bibinfo {author} {\bibfnamefont {J.~D.}\ \bibnamefont
  {Joannopoulos}}, \ and\ \bibinfo {author} {\bibfnamefont {M.}~\bibnamefont
  {Solja{\v{c}}i{\'c}}},\ }\href@noop {} {\bibfield  {journal} {\bibinfo
  {journal} {Nature Photonics}\ }\textbf {\bibinfo {volume} {8}},\ \bibinfo
  {pages} {821} (\bibinfo {year} {2014})}\BibitemShut {NoStop}%
\bibitem [{\citenamefont {Ozawa}\ \emph {et~al.}(2019)\citenamefont {Ozawa},
  \citenamefont {Price}, \citenamefont {Amo}, \citenamefont {Goldman},
  \citenamefont {Hafezi}, \citenamefont {Lu}, \citenamefont {Rechtsman},
  \citenamefont {Schuster}, \citenamefont {Simon}, \citenamefont {Zilberberg}
  \emph {et~al.}}]{ozawa2019topological}%
  \BibitemOpen
  \bibfield  {author} {\bibinfo {author} {\bibfnamefont {T.}~\bibnamefont
  {Ozawa}}, \bibinfo {author} {\bibfnamefont {H.~M.}\ \bibnamefont {Price}},
  \bibinfo {author} {\bibfnamefont {A.}~\bibnamefont {Amo}}, \bibinfo {author}
  {\bibfnamefont {N.}~\bibnamefont {Goldman}}, \bibinfo {author} {\bibfnamefont
  {M.}~\bibnamefont {Hafezi}}, \bibinfo {author} {\bibfnamefont
  {L.}~\bibnamefont {Lu}}, \bibinfo {author} {\bibfnamefont {M.~C.}\
  \bibnamefont {Rechtsman}}, \bibinfo {author} {\bibfnamefont {D.}~\bibnamefont
  {Schuster}}, \bibinfo {author} {\bibfnamefont {J.}~\bibnamefont {Simon}},
  \bibinfo {author} {\bibfnamefont {O.}~\bibnamefont {Zilberberg}},  \emph
  {et~al.},\ }\href@noop {} {\bibfield  {journal} {\bibinfo  {journal} {Reviews
  of Modern Physics}\ }\textbf {\bibinfo {volume} {91}},\ \bibinfo {pages}
  {015006} (\bibinfo {year} {2019})}\BibitemShut {NoStop}%
\bibitem [{\citenamefont {Barthelemy}\ \emph {et~al.}(1985)\citenamefont
  {Barthelemy}, \citenamefont {Maneuf},\ and\ \citenamefont
  {Froehly}}]{barthelemy1985propagation}%
  \BibitemOpen
  \bibfield  {author} {\bibinfo {author} {\bibfnamefont {A.}~\bibnamefont
  {Barthelemy}}, \bibinfo {author} {\bibfnamefont {S.}~\bibnamefont {Maneuf}},
  \ and\ \bibinfo {author} {\bibfnamefont {C.}~\bibnamefont {Froehly}},\
  }\href@noop {} {\bibfield  {journal} {\bibinfo  {journal} {Optics
  Communications}\ }\textbf {\bibinfo {volume} {55}},\ \bibinfo {pages} {201}
  (\bibinfo {year} {1985})}\BibitemShut {NoStop}%
\bibitem [{\citenamefont {Christodoulides}\ and\ \citenamefont
  {Joseph}(1988)}]{christodoulides1988discrete}%
  \BibitemOpen
  \bibfield  {author} {\bibinfo {author} {\bibfnamefont {D.}~\bibnamefont
  {Christodoulides}}\ and\ \bibinfo {author} {\bibfnamefont {R.}~\bibnamefont
  {Joseph}},\ }\href@noop {} {\bibfield  {journal} {\bibinfo  {journal} {Optics
  Letters}\ }\textbf {\bibinfo {volume} {13}},\ \bibinfo {pages} {794}
  (\bibinfo {year} {1988})}\BibitemShut {NoStop}%
\bibitem [{\citenamefont {Segev}\ \emph {et~al.}(1992)\citenamefont {Segev},
  \citenamefont {Crosignani}, \citenamefont {Yariv},\ and\ \citenamefont
  {Fischer}}]{segev1992spatial}%
  \BibitemOpen
  \bibfield  {author} {\bibinfo {author} {\bibfnamefont {M.}~\bibnamefont
  {Segev}}, \bibinfo {author} {\bibfnamefont {B.}~\bibnamefont {Crosignani}},
  \bibinfo {author} {\bibfnamefont {A.}~\bibnamefont {Yariv}}, \ and\ \bibinfo
  {author} {\bibfnamefont {B.}~\bibnamefont {Fischer}},\ }\href@noop {}
  {\bibfield  {journal} {\bibinfo  {journal} {Physical Review Letters}\
  }\textbf {\bibinfo {volume} {68}},\ \bibinfo {pages} {923} (\bibinfo {year}
  {1992})}\BibitemShut {NoStop}%
\bibitem [{\citenamefont {Eisenberg}\ \emph {et~al.}(1998)\citenamefont
  {Eisenberg}, \citenamefont {Silberberg}, \citenamefont {Morandotti},
  \citenamefont {Boyd},\ and\ \citenamefont
  {Aitchison}}]{eisenberg1998discrete}%
  \BibitemOpen
  \bibfield  {author} {\bibinfo {author} {\bibfnamefont {H.}~\bibnamefont
  {Eisenberg}}, \bibinfo {author} {\bibfnamefont {Y.}~\bibnamefont
  {Silberberg}}, \bibinfo {author} {\bibfnamefont {R.}~\bibnamefont
  {Morandotti}}, \bibinfo {author} {\bibfnamefont {A.}~\bibnamefont {Boyd}}, \
  and\ \bibinfo {author} {\bibfnamefont {J.}~\bibnamefont {Aitchison}},\
  }\href@noop {} {\bibfield  {journal} {\bibinfo  {journal} {Physical Review
  Letters}\ }\textbf {\bibinfo {volume} {81}},\ \bibinfo {pages} {3383}
  (\bibinfo {year} {1998})}\BibitemShut {NoStop}%
\bibitem [{\citenamefont {Stegeman}\ and\ \citenamefont
  {Segev}(1999)}]{stegeman1999optical}%
  \BibitemOpen
  \bibfield  {author} {\bibinfo {author} {\bibfnamefont {G.~I.}\ \bibnamefont
  {Stegeman}}\ and\ \bibinfo {author} {\bibfnamefont {M.}~\bibnamefont
  {Segev}},\ }\href@noop {} {\bibfield  {journal} {\bibinfo  {journal}
  {Science}\ }\textbf {\bibinfo {volume} {286}},\ \bibinfo {pages} {1518}
  (\bibinfo {year} {1999})}\BibitemShut {NoStop}%
\bibitem [{\citenamefont {Fleischer}\ \emph {et~al.}(2003)\citenamefont
  {Fleischer}, \citenamefont {Segev}, \citenamefont {Efremidis},\ and\
  \citenamefont {Christodoulides}}]{fleischer2003observation}%
  \BibitemOpen
  \bibfield  {author} {\bibinfo {author} {\bibfnamefont {J.~W.}\ \bibnamefont
  {Fleischer}}, \bibinfo {author} {\bibfnamefont {M.}~\bibnamefont {Segev}},
  \bibinfo {author} {\bibfnamefont {N.~K.}\ \bibnamefont {Efremidis}}, \ and\
  \bibinfo {author} {\bibfnamefont {D.~N.}\ \bibnamefont {Christodoulides}},\
  }\href@noop {} {\bibfield  {journal} {\bibinfo  {journal} {Nature}\ }\textbf
  {\bibinfo {volume} {422}},\ \bibinfo {pages} {147} (\bibinfo {year}
  {2003})}\BibitemShut {NoStop}%
\bibitem [{\citenamefont {Mukherjee}\ \emph {et~al.}(2017)\citenamefont
  {Mukherjee}, \citenamefont {Spracklen}, \citenamefont {Valiente},
  \citenamefont {Andersson}, \citenamefont {{\"O}hberg}, \citenamefont
  {Goldman},\ and\ \citenamefont {Thomson}}]{mukherjee2017experimental}%
  \BibitemOpen
  \bibfield  {author} {\bibinfo {author} {\bibfnamefont {S.}~\bibnamefont
  {Mukherjee}}, \bibinfo {author} {\bibfnamefont {A.}~\bibnamefont
  {Spracklen}}, \bibinfo {author} {\bibfnamefont {M.}~\bibnamefont {Valiente}},
  \bibinfo {author} {\bibfnamefont {E.}~\bibnamefont {Andersson}}, \bibinfo
  {author} {\bibfnamefont {P.}~\bibnamefont {{\"O}hberg}}, \bibinfo {author}
  {\bibfnamefont {N.}~\bibnamefont {Goldman}}, \ and\ \bibinfo {author}
  {\bibfnamefont {R.~R.}\ \bibnamefont {Thomson}},\ }\href@noop {} {\bibfield
  {journal} {\bibinfo  {journal} {Nature Communications}\ }\textbf {\bibinfo
  {volume} {8}},\ \bibinfo {pages} {13918} (\bibinfo {year}
  {2017})}\BibitemShut {NoStop}%
\bibitem [{\citenamefont {Maczewsky}\ \emph {et~al.}(2017)\citenamefont
  {Maczewsky}, \citenamefont {Zeuner}, \citenamefont {Nolte},\ and\
  \citenamefont {Szameit}}]{maczewsky2017observation}%
  \BibitemOpen
  \bibfield  {author} {\bibinfo {author} {\bibfnamefont {L.~J.}\ \bibnamefont
  {Maczewsky}}, \bibinfo {author} {\bibfnamefont {J.~M.}\ \bibnamefont
  {Zeuner}}, \bibinfo {author} {\bibfnamefont {S.}~\bibnamefont {Nolte}}, \
  and\ \bibinfo {author} {\bibfnamefont {A.}~\bibnamefont {Szameit}},\
  }\href@noop {} {\bibfield  {journal} {\bibinfo  {journal} {Nature
  Communications}\ }\textbf {\bibinfo {volume} {8}},\ \bibinfo {pages} {13756}
  (\bibinfo {year} {2017})}\BibitemShut {NoStop}%
\bibitem [{\citenamefont {Lumer}\ \emph {et~al.}(2013)\citenamefont {Lumer},
  \citenamefont {Plotnik}, \citenamefont {Rechtsman},\ and\ \citenamefont
  {Segev}}]{lumer2013self}%
  \BibitemOpen
  \bibfield  {author} {\bibinfo {author} {\bibfnamefont {Y.}~\bibnamefont
  {Lumer}}, \bibinfo {author} {\bibfnamefont {Y.}~\bibnamefont {Plotnik}},
  \bibinfo {author} {\bibfnamefont {M.~C.}\ \bibnamefont {Rechtsman}}, \ and\
  \bibinfo {author} {\bibfnamefont {M.}~\bibnamefont {Segev}},\ }\href@noop {}
  {\bibfield  {journal} {\bibinfo  {journal} {Physical Review Letters}\
  }\textbf {\bibinfo {volume} {111}},\ \bibinfo {pages} {243905} (\bibinfo
  {year} {2013})}\BibitemShut {NoStop}%
\bibitem [{\citenamefont {Ablowitz}\ \emph {et~al.}(2014)\citenamefont
  {Ablowitz}, \citenamefont {Curtis},\ and\ \citenamefont
  {Ma}}]{ablowitz2014linear}%
  \BibitemOpen
  \bibfield  {author} {\bibinfo {author} {\bibfnamefont {M.~J.}\ \bibnamefont
  {Ablowitz}}, \bibinfo {author} {\bibfnamefont {C.~W.}\ \bibnamefont
  {Curtis}}, \ and\ \bibinfo {author} {\bibfnamefont {Y.-P.}\ \bibnamefont
  {Ma}},\ }\href@noop {} {\bibfield  {journal} {\bibinfo  {journal} {Physical
  Review A}\ }\textbf {\bibinfo {volume} {90}},\ \bibinfo {pages} {023813}
  (\bibinfo {year} {2014})}\BibitemShut {NoStop}%
\bibitem [{\citenamefont {Leykam}\ and\ \citenamefont
  {Chong}(2016)}]{leykam2016edge}%
  \BibitemOpen
  \bibfield  {author} {\bibinfo {author} {\bibfnamefont {D.}~\bibnamefont
  {Leykam}}\ and\ \bibinfo {author} {\bibfnamefont {Y.~D.}\ \bibnamefont
  {Chong}},\ }\href@noop {} {\bibfield  {journal} {\bibinfo  {journal}
  {Physical Review Letters}\ }\textbf {\bibinfo {volume} {117}},\ \bibinfo
  {pages} {143901} (\bibinfo {year} {2016})}\BibitemShut {NoStop}%
\bibitem [{\citenamefont {Marzuola}\ \emph {et~al.}(2019)\citenamefont
  {Marzuola}, \citenamefont {Rechtsman}, \citenamefont {Osting},\ and\
  \citenamefont {Bandres}}]{marzuola2019bulk}%
  \BibitemOpen
  \bibfield  {author} {\bibinfo {author} {\bibfnamefont {J.~L.}\ \bibnamefont
  {Marzuola}}, \bibinfo {author} {\bibfnamefont {M.}~\bibnamefont {Rechtsman}},
  \bibinfo {author} {\bibfnamefont {B.}~\bibnamefont {Osting}}, \ and\ \bibinfo
  {author} {\bibfnamefont {M.}~\bibnamefont {Bandres}},\ }\href@noop {}
  {\bibfield  {journal} {\bibinfo  {journal} {arXiv preprint arXiv:1904.10312}\
  } (\bibinfo {year} {2019})}\BibitemShut {NoStop}%
\bibitem [{\citenamefont {Mukherjee}\ \emph {et~al.}(2018)\citenamefont
  {Mukherjee}, \citenamefont {Chandrasekharan}, \citenamefont {{\"O}hberg},
  \citenamefont {Goldman},\ and\ \citenamefont {Thomson}}]{mukherjee2018state}%
  \BibitemOpen
  \bibfield  {author} {\bibinfo {author} {\bibfnamefont {S.}~\bibnamefont
  {Mukherjee}}, \bibinfo {author} {\bibfnamefont {H.~K.}\ \bibnamefont
  {Chandrasekharan}}, \bibinfo {author} {\bibfnamefont {P.}~\bibnamefont
  {{\"O}hberg}}, \bibinfo {author} {\bibfnamefont {N.}~\bibnamefont {Goldman}},
  \ and\ \bibinfo {author} {\bibfnamefont {R.~R.}\ \bibnamefont {Thomson}},\
  }\href@noop {} {\bibfield  {journal} {\bibinfo  {journal} {Nature
  Communications}\ }\textbf {\bibinfo {volume} {9}},\ \bibinfo {pages} {4209}
  (\bibinfo {year} {2018})}\BibitemShut {NoStop}%
\bibitem [{\citenamefont {Kitagawa}\ \emph {et~al.}(2010)\citenamefont
  {Kitagawa}, \citenamefont {Berg}, \citenamefont {Rudner},\ and\ \citenamefont
  {Demler}}]{kitagawa2010topological}%
  \BibitemOpen
  \bibfield  {author} {\bibinfo {author} {\bibfnamefont {T.}~\bibnamefont
  {Kitagawa}}, \bibinfo {author} {\bibfnamefont {E.}~\bibnamefont {Berg}},
  \bibinfo {author} {\bibfnamefont {M.}~\bibnamefont {Rudner}}, \ and\ \bibinfo
  {author} {\bibfnamefont {E.}~\bibnamefont {Demler}},\ }\href@noop {}
  {\bibfield  {journal} {\bibinfo  {journal} {Physical Review B}\ }\textbf
  {\bibinfo {volume} {82}},\ \bibinfo {pages} {235114} (\bibinfo {year}
  {2010})}\BibitemShut {NoStop}%
\bibitem [{top()}]{toposolAnimations}%
  \BibitemOpen
  \href {https://leptos.psu.edu/2019/11/01/supp-anim/} {}\bibinfo {note} {See
  Supplemental Animations at
  {\color{blue}https://leptos.psu.edu/2019/11/01/supp-anim/}}\BibitemShut
  {NoStop}%
\bibitem [{\citenamefont {Szameit}\ \emph {et~al.}(2006)\citenamefont
  {Szameit}, \citenamefont {Burghoff}, \citenamefont {Pertsch}, \citenamefont
  {Nolte}, \citenamefont {T{\"u}nnermann},\ and\ \citenamefont
  {Lederer}}]{szameit2006two}%
  \BibitemOpen
  \bibfield  {author} {\bibinfo {author} {\bibfnamefont {A.}~\bibnamefont
  {Szameit}}, \bibinfo {author} {\bibfnamefont {J.}~\bibnamefont {Burghoff}},
  \bibinfo {author} {\bibfnamefont {T.}~\bibnamefont {Pertsch}}, \bibinfo
  {author} {\bibfnamefont {S.}~\bibnamefont {Nolte}}, \bibinfo {author}
  {\bibfnamefont {A.}~\bibnamefont {T{\"u}nnermann}}, \ and\ \bibinfo {author}
  {\bibfnamefont {F.}~\bibnamefont {Lederer}},\ }\href@noop {} {\bibfield
  {journal} {\bibinfo  {journal} {Optics Express}\ }\textbf {\bibinfo {volume}
  {14}},\ \bibinfo {pages} {6055} (\bibinfo {year} {2006})}\BibitemShut
  {NoStop}%
\bibitem [{\citenamefont {Cao}\ \emph {et~al.}(2018)\citenamefont {Cao},
  \citenamefont {Fatemi}, \citenamefont {Fang}, \citenamefont {Watanabe},
  \citenamefont {Taniguchi}, \citenamefont {Kaxiras},\ and\ \citenamefont
  {Jarillo-Herrero}}]{cao2018unconventional}%
  \BibitemOpen
  \bibfield  {author} {\bibinfo {author} {\bibfnamefont {Y.}~\bibnamefont
  {Cao}}, \bibinfo {author} {\bibfnamefont {V.}~\bibnamefont {Fatemi}},
  \bibinfo {author} {\bibfnamefont {S.}~\bibnamefont {Fang}}, \bibinfo {author}
  {\bibfnamefont {K.}~\bibnamefont {Watanabe}}, \bibinfo {author}
  {\bibfnamefont {T.}~\bibnamefont {Taniguchi}}, \bibinfo {author}
  {\bibfnamefont {E.}~\bibnamefont {Kaxiras}}, \ and\ \bibinfo {author}
  {\bibfnamefont {P.}~\bibnamefont {Jarillo-Herrero}},\ }\href@noop {}
  {\bibfield  {journal} {\bibinfo  {journal} {Nature}\ }\textbf {\bibinfo
  {volume} {556}},\ \bibinfo {pages} {43} (\bibinfo {year} {2018})}\BibitemShut
  {NoStop}%
\bibitem [{\citenamefont {Clark}\ \emph {et~al.}(2019)\citenamefont {Clark},
  \citenamefont {Schine}, \citenamefont {Baum}, \citenamefont {Jia},\ and\
  \citenamefont {Simon}}]{clark2019observation}%
  \BibitemOpen
  \bibfield  {author} {\bibinfo {author} {\bibfnamefont {L.~W.}\ \bibnamefont
  {Clark}}, \bibinfo {author} {\bibfnamefont {N.}~\bibnamefont {Schine}},
  \bibinfo {author} {\bibfnamefont {C.}~\bibnamefont {Baum}}, \bibinfo {author}
  {\bibfnamefont {N.}~\bibnamefont {Jia}}, \ and\ \bibinfo {author}
  {\bibfnamefont {J.}~\bibnamefont {Simon}},\ }\href@noop {} {\bibfield
  {journal} {\bibinfo  {journal} {arXiv preprint arXiv:1907.05872}\ } (\bibinfo
  {year} {2019})}\BibitemShut {NoStop}%
\bibitem [{\citenamefont {Hadad}\ \emph {et~al.}(2016)\citenamefont {Hadad},
  \citenamefont {Khanikaev},\ and\ \citenamefont {Alu}}]{hadad2016self}%
  \BibitemOpen
  \bibfield  {author} {\bibinfo {author} {\bibfnamefont {Y.}~\bibnamefont
  {Hadad}}, \bibinfo {author} {\bibfnamefont {A.~B.}\ \bibnamefont
  {Khanikaev}}, \ and\ \bibinfo {author} {\bibfnamefont {A.}~\bibnamefont
  {Alu}},\ }\href@noop {} {\bibfield  {journal} {\bibinfo  {journal} {Physical
  Review B}\ }\textbf {\bibinfo {volume} {93}},\ \bibinfo {pages} {155112}
  (\bibinfo {year} {2016})}\BibitemShut {NoStop}%
\bibitem [{\citenamefont {Leykam}\ \emph {et~al.}(2018)\citenamefont {Leykam},
  \citenamefont {Mittal}, \citenamefont {Hafezi},\ and\ \citenamefont
  {Chong}}]{PhysRevLett.121.023901}%
  \BibitemOpen
  \bibfield  {author} {\bibinfo {author} {\bibfnamefont {D.}~\bibnamefont
  {Leykam}}, \bibinfo {author} {\bibfnamefont {S.}~\bibnamefont {Mittal}},
  \bibinfo {author} {\bibfnamefont {M.}~\bibnamefont {Hafezi}}, \ and\ \bibinfo
  {author} {\bibfnamefont {Y.~D.}\ \bibnamefont {Chong}},\ }\href@noop {}
  {\bibfield  {journal} {\bibinfo  {journal} {Physical Review Letters}\
  }\textbf {\bibinfo {volume} {121}},\ \bibinfo {pages} {023901} (\bibinfo
  {year} {2018})}\BibitemShut {NoStop}%
\bibitem [{\citenamefont {He}\ \emph {et~al.}(2019)\citenamefont {He},
  \citenamefont {Addison}, \citenamefont {Jin}, \citenamefont {Mele},
  \citenamefont {Johnson},\ and\ \citenamefont {Zhen}}]{he_floquet_2019}%
  \BibitemOpen
  \bibfield  {author} {\bibinfo {author} {\bibfnamefont {L.}~\bibnamefont
  {He}}, \bibinfo {author} {\bibfnamefont {Z.}~\bibnamefont {Addison}},
  \bibinfo {author} {\bibfnamefont {J.}~\bibnamefont {Jin}}, \bibinfo {author}
  {\bibfnamefont {E.~J.}\ \bibnamefont {Mele}}, \bibinfo {author}
  {\bibfnamefont {S.~G.}\ \bibnamefont {Johnson}}, \ and\ \bibinfo {author}
  {\bibfnamefont {B.}~\bibnamefont {Zhen}},\ }\href@noop {} {\bibfield
  {journal} {\bibinfo  {journal} {Nature Communications}\ }\textbf {\bibinfo
  {volume} {10}},\ \bibinfo {pages} {4194} (\bibinfo {year}
  {2019})}\BibitemShut {NoStop}%
\bibitem [{\citenamefont {Kruk}\ \emph {et~al.}(2019)\citenamefont {Kruk},
  \citenamefont {Poddubny}, \citenamefont {Smirnova}, \citenamefont {Wang},
  \citenamefont {Slobozhanyuk}, \citenamefont {Shorokhov}, \citenamefont
  {Kravchenko}, \citenamefont {Luther-Davies},\ and\ \citenamefont
  {Kivshar}}]{kruk2019nonlinear}%
  \BibitemOpen
  \bibfield  {author} {\bibinfo {author} {\bibfnamefont {S.}~\bibnamefont
  {Kruk}}, \bibinfo {author} {\bibfnamefont {A.}~\bibnamefont {Poddubny}},
  \bibinfo {author} {\bibfnamefont {D.}~\bibnamefont {Smirnova}}, \bibinfo
  {author} {\bibfnamefont {L.}~\bibnamefont {Wang}}, \bibinfo {author}
  {\bibfnamefont {A.}~\bibnamefont {Slobozhanyuk}}, \bibinfo {author}
  {\bibfnamefont {A.}~\bibnamefont {Shorokhov}}, \bibinfo {author}
  {\bibfnamefont {I.}~\bibnamefont {Kravchenko}}, \bibinfo {author}
  {\bibfnamefont {B.}~\bibnamefont {Luther-Davies}}, \ and\ \bibinfo {author}
  {\bibfnamefont {Y.}~\bibnamefont {Kivshar}},\ }\href@noop {} {\bibfield
  {journal} {\bibinfo  {journal} {Nature Nanotechnology}\ }\textbf {\bibinfo
  {volume} {14}},\ \bibinfo {pages} {126} (\bibinfo {year} {2019})}\BibitemShut
  {NoStop}%
\bibitem [{\citenamefont {Smirnova}\ \emph {et~al.}(2019)\citenamefont
  {Smirnova}, \citenamefont {Kruk}, \citenamefont {Leykam}, \citenamefont
  {Melik-Gaykazyan}, \citenamefont {Choi},\ and\ \citenamefont
  {Kivshar}}]{PhysRevLett.123.103901}%
  \BibitemOpen
  \bibfield  {author} {\bibinfo {author} {\bibfnamefont {D.}~\bibnamefont
  {Smirnova}}, \bibinfo {author} {\bibfnamefont {S.}~\bibnamefont {Kruk}},
  \bibinfo {author} {\bibfnamefont {D.}~\bibnamefont {Leykam}}, \bibinfo
  {author} {\bibfnamefont {E.}~\bibnamefont {Melik-Gaykazyan}}, \bibinfo
  {author} {\bibfnamefont {D.-Y.}\ \bibnamefont {Choi}}, \ and\ \bibinfo
  {author} {\bibfnamefont {Y.}~\bibnamefont {Kivshar}},\ }\href@noop {}
  {\bibfield  {journal} {\bibinfo  {journal} {Physical Review Letters}\
  }\textbf {\bibinfo {volume} {123}},\ \bibinfo {pages} {103901} (\bibinfo
  {year} {2019})}\BibitemShut {NoStop}%
\bibitem [{\citenamefont {Davis}\ \emph {et~al.}(1996)\citenamefont {Davis},
  \citenamefont {Miura}, \citenamefont {Sugimoto},\ and\ \citenamefont
  {Hirao}}]{davis1996writing}%
  \BibitemOpen
  \bibfield  {author} {\bibinfo {author} {\bibfnamefont {K.~M.}\ \bibnamefont
  {Davis}}, \bibinfo {author} {\bibfnamefont {K.}~\bibnamefont {Miura}},
  \bibinfo {author} {\bibfnamefont {N.}~\bibnamefont {Sugimoto}}, \ and\
  \bibinfo {author} {\bibfnamefont {K.}~\bibnamefont {Hirao}},\ }\href@noop {}
  {\bibfield  {journal} {\bibinfo  {journal} {Optics Letters}\ }\textbf
  {\bibinfo {volume} {21}},\ \bibinfo {pages} {1729} (\bibinfo {year}
  {1996})}\BibitemShut {NoStop}%
\bibitem [{\citenamefont {Szameit}\ and\ \citenamefont
  {Nolte}(2010)}]{szameit2010discrete}%
  \BibitemOpen
  \bibfield  {author} {\bibinfo {author} {\bibfnamefont {A.}~\bibnamefont
  {Szameit}}\ and\ \bibinfo {author} {\bibfnamefont {S.}~\bibnamefont
  {Nolte}},\ }\href@noop {} {\bibfield  {journal} {\bibinfo  {journal} {Journal
  of Physics B: Atomic, Molecular and Optical Physics}\ }\textbf {\bibinfo
  {volume} {43}},\ \bibinfo {pages} {163001} (\bibinfo {year}
  {2010})}\BibitemShut {NoStop}%
\bibitem [{\citenamefont {Ams}\ \emph {et~al.}(2005)\citenamefont {Ams},
  \citenamefont {Marshall}, \citenamefont {Spence},\ and\ \citenamefont
  {Withford}}]{ams2005slit}%
  \BibitemOpen
  \bibfield  {author} {\bibinfo {author} {\bibfnamefont {M.}~\bibnamefont
  {Ams}}, \bibinfo {author} {\bibfnamefont {G.}~\bibnamefont {Marshall}},
  \bibinfo {author} {\bibfnamefont {D.}~\bibnamefont {Spence}}, \ and\ \bibinfo
  {author} {\bibfnamefont {M.}~\bibnamefont {Withford}},\ }\href@noop {}
  {\bibfield  {journal} {\bibinfo  {journal} {Optics Express}\ }\textbf
  {\bibinfo {volume} {13}},\ \bibinfo {pages} {5676} (\bibinfo {year}
  {2005})}\BibitemShut {NoStop}%
\bibitem [{\citenamefont {Sheik-Bahae}\ \emph {et~al.}(1990)\citenamefont
  {Sheik-Bahae}, \citenamefont {Said}, \citenamefont {Wei}, \citenamefont
  {Hagan},\ and\ \citenamefont {Van~Stryland}}]{sheik1990sensitive}%
  \BibitemOpen
  \bibfield  {author} {\bibinfo {author} {\bibfnamefont {M.}~\bibnamefont
  {Sheik-Bahae}}, \bibinfo {author} {\bibfnamefont {A.~A.}\ \bibnamefont
  {Said}}, \bibinfo {author} {\bibfnamefont {T.-H.}\ \bibnamefont {Wei}},
  \bibinfo {author} {\bibfnamefont {D.~J.}\ \bibnamefont {Hagan}}, \ and\
  \bibinfo {author} {\bibfnamefont {E.~W.}\ \bibnamefont {Van~Stryland}},\
  }\href@noop {} {\bibfield  {journal} {\bibinfo  {journal} {IEEE Journal of
  Quantum Electronics}\ }\textbf {\bibinfo {volume} {26}},\ \bibinfo {pages}
  {760} (\bibinfo {year} {1990})}\BibitemShut {NoStop}%
\bibitem [{\citenamefont {Jensen}(1982)}]{jensen1982nonlinear}%
  \BibitemOpen
  \bibfield  {author} {\bibinfo {author} {\bibfnamefont {S.}~\bibnamefont
  {Jensen}},\ }\href@noop {} {\bibfield  {journal} {\bibinfo  {journal} {IEEE
  Journal of Quantum Electronics}\ }\textbf {\bibinfo {volume} {18}},\ \bibinfo
  {pages} {1580} (\bibinfo {year} {1982})}\BibitemShut {NoStop}%
\end{thebibliography}

%


\clearpage

\onecolumngrid
\appendix

\vspace{1.0cm}

\section{\large {Supplementary Information} \vspace*{0.3cm}}

\twocolumngrid

In the following sections of the supplementary information, we present experimental details on waveguide fabrication, systematic nonlinear characterisation of the photonic devices, the observation of topologically trivial bulk solitons in a static square lattice and finally, provide numerical results associated with the topological solitons.

\begin{figure*}[hbt]
\center
\includegraphics[width=15cm]{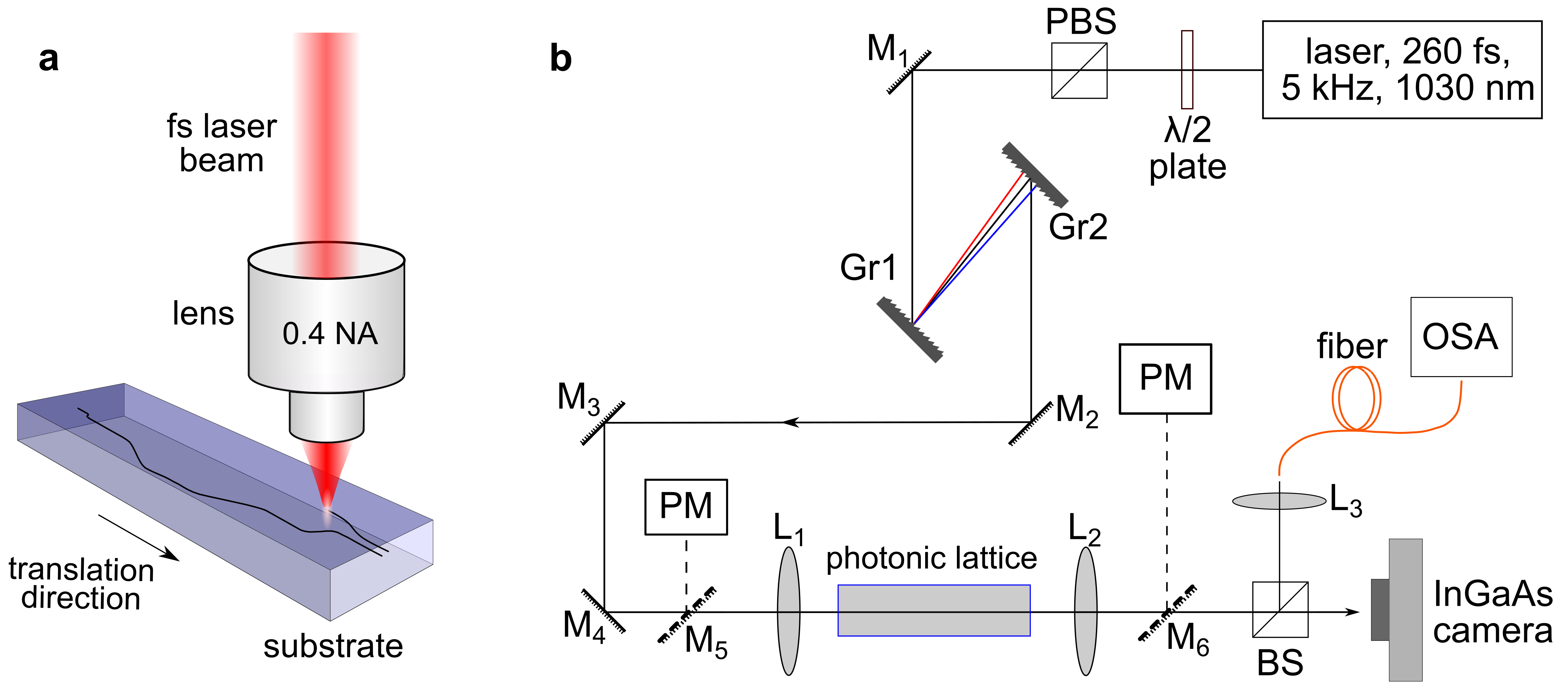}
\caption{{\bf Fabrication and characterisation setups. a,} Simplified sketch of the femtosecond-laser-writing process. To fabricate each waveguide with a three-dimensional path, the substrate is translated once through the focus of the laser pulses with optimised fabrication parameters. {\bf b,} Schematic diagram of the characterisation setup. Here, M$_{1-6}$ are silver-coated mirrors, M$_{5, 6}$ are on flip mounts, L$_{1-3}$ are air-coated convex lenses, BS is a $50\!:\!50$ beam splitter and PM is a power meter. Optical pulse trains (at $1030$~nm wavelength, $5$~kHz pulse repetition rate and $260$~fs pulse duration) are generated using a Yb-doped fibre laser (Menlo BlueCut) system. Optical power is controlled using a half-wave plate and a polarising beam splitter (PBS). Optical pulses are temporally stretched, as well as down-chirped, using a pair of parallel gratings. Intensity distribution at the output of the photonic lattice is measured using an InGaAs camera and the optical spectrum is measured using an optical spectrum analyser (OSA). }
\label{Supp_fig1}
\end{figure*}

\section{1. Fabrication} \label{Fab}
The optical waveguide arrays were fabricated using femtosecond laser writing~\cite{davis1996writing, szameit2010discrete, mukherjee2017experimental} (see Fig~\ref{Supp_fig1}a). Circularly-polarised sub-picosecond ($260$~fs) pulse trains at $1030$~nm wavelength and $500$~kHz repetition rate were generated using a commercially available Yb-doped fibre laser (Menlo BlueCut) system. Using an aspheric lens with $0.4$ numerical aperture, the laser beam was focused inside a borosilicate glass (Corning Eagle XG) sample 
mounted on high-precision $x$-$y$-$z$ translation stages (Aerotech). To fabricate each waveguide, the substrate was translated once through the focus of the laser beam at $6$~mm/s speed. We used a slit-beam shaping (not shown in Fig.~\ref{Supp_fig1}a) 
technique~\cite{ams2005slit, mukherjee2017experimental} to control the cross-sectional refractive index profile of the waveguides. By controlling the fabrication parameters, we can achieve a positive refractive index modification that ranges between approximately $10^{-4}$ to $10^{-3}$.
Pulse energy and translation speed of fabrication were optimised to achieve tightly confined waveguides with low optical losses. 

Straight single-mode waveguides at $1030 \pm 4$~nm wavelength exhibited $\approx\!0.5$~dB/cm propagation loss. The total bend loss of the modulated waveguides forming the topological photonic lattice was estimated to be $\approx\!4$ dB, through the entire sample.
Our waveguides exhibited neither significant polarisation-dependent losses nor rotation of polarization during $76$~mm propagation. 
The evanescent coupling strength was measured by characterising a set of two-waveguide directional couplers. We observed an exponential variation of coupling strength with inter-waveguide spacing, as would be expected. 
A white-light transmission micrograph of the lattice (cross-section) is shown in Fig.~\ref{fig1}d. Each waveguide in the lattice supports an elliptical fundamental mode near $1030$~nm wavelength with
$8\pm0.2 \; \mu$m and $10.8\pm0.2 \; \mu$m
mode field diameters (the diameter at which intensity is $1/e^2$ of its maximum value) along the horizontal and vertical directions, respectively. 
It is expected that the evanescent coupling $J$ will have small fluctuation due to fabrication issues such as  random fluctuation in inter-waveguide spacing. The off-diagonal disorder i.e.,~the variation of $J$ was estimated to be $\approx\!5$\,\% of its mean value. 
During fabrication, the diagonal disorder can primarily occur (in our case) due to the depth-dependent aberrations of the laser beam by the air-glass interface. 
However, our coupler characterisations demonstrated the near-complete transfer of optical power from the input waveguide to the nearest waveguide after one coupling length [$ \equiv \! \pi/(2J)$]. A complete transfer of power in an evanescently coupled directional coupler can only happen if the waveguides support modes with identical effective refractive indices. Hence, we can conclude that the local variation of modal effective refractive indices (i.e.,~diagonal disorder) is negligibly small.

\begin{figure*}[t!]
\center
\includegraphics[width=\textwidth]{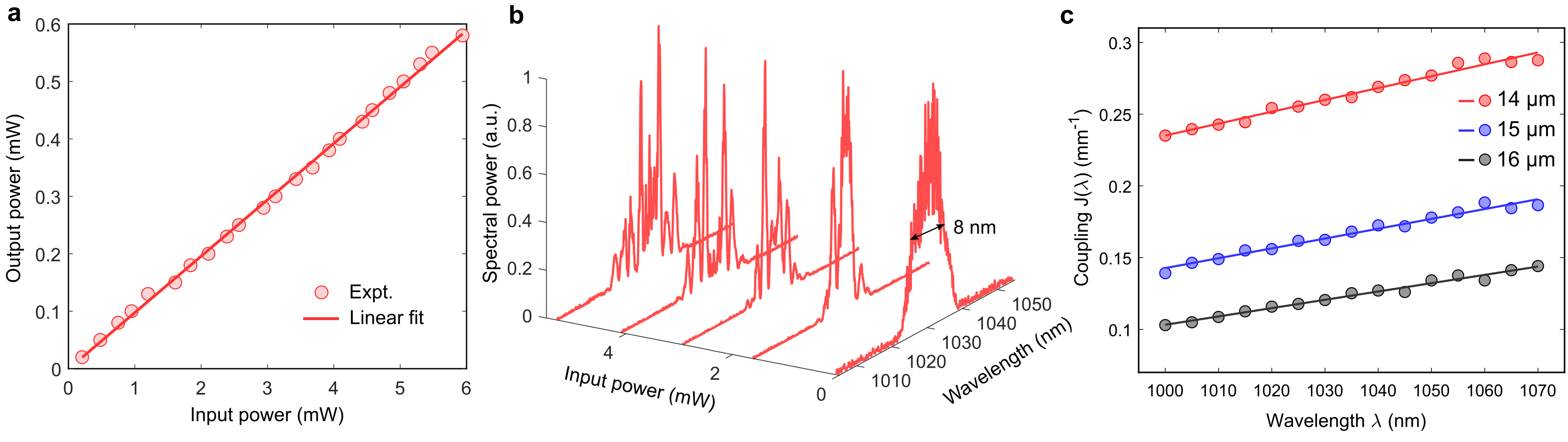}
\caption{{\bf Negligible effects due to multi-photon absorption and self-phase modulation. a,} For each nonlinear experiment, both input and output power were monitored. The linear variation of average output power with average input power implies that the nonlinear loss due to multi-photon absorption is insignificant. {\bf b,} Self-phase modulation (SPM): normalised spectral power as a function of average input power. The spectral width in the linear regime is $\approx 8$ nm (FWHM). Due to SPM the spectrum changes at higher powers; however, the maximal spectral width after $76$~mm of propagation was $<20$ nm for the maximal required nonlinearity. 
{\bf c,} Measured variation of evanescent coupling strength $J(d, \lambda)$ as a function of wavelength $\lambda$ for three different inter-waveguide separations, $d\!=\!14, 15$ and $16\, \mu$m; the solid lines are linear fits. For $1030\pm 10$~nm wavelength range, $J$ varies by $\Delta J/J\!\approx\! \pm4$~\%.
}
\label{Supp_fig2}
\end{figure*}

\section{2. Characterisation}  \label{Char}
A simplified schematic of the characterisation setup 
is shown in  Fig~\ref{Supp_fig1}b. 
Optical pulses from the Menlo BlueCut laser system were temporally stretched to $t_p\!\approx\!2$~ps using a pair of parallel gratings which introduce a negative group velocity dispersion. As a result of temporal stretching, pulses were simultaneously down-chirped. The photonic lattice was mounted on four-axis stages for precise angular and translational alignment. Using free-space input coupling, we can launch light into a desired waveguide of the photonic lattice. Intensity distributions at the output of the photonic lattice can then be imaged on an InGaAs camera (ICI systems). We note that our measurements are not affected by the wavelength-dependent quantum efficiency (QE) since the QE of this camera is flat near the wavelength range of interest. 

We tuned input power using a power controller consisting of a half-wave plate and a polarising beam splitter (PBS). For each measurement both input and output powers were monitored and it was found that the output power varies linearly (see Fig.~\ref{Supp_fig2}a) with input power implying that the nonlinear loss due to multi-photon absorption is insignificant. 
Additionally, we monitored the spectral broadening due to self-phase modulation (SPM) for each measurement using an optical spectrum analyser (Anritsu MS9740A).
Fig.~\ref{Supp_fig2}b presents the variation of normalised spectral power as a function of input power. 
The spectral width at low power (linear regime) is $\approx 8$~nm (FWHM). As nonlinearity was increased by increasing the input power, the spectrum started to change. We optimised the grating separation such that the maximal spectral width after $76$~mm propagation is $<\!20$~nm for the maximal nonlinearity required for our experiments.

\begin{figure*}[htb]
\center
\includegraphics[width=12.6cm]{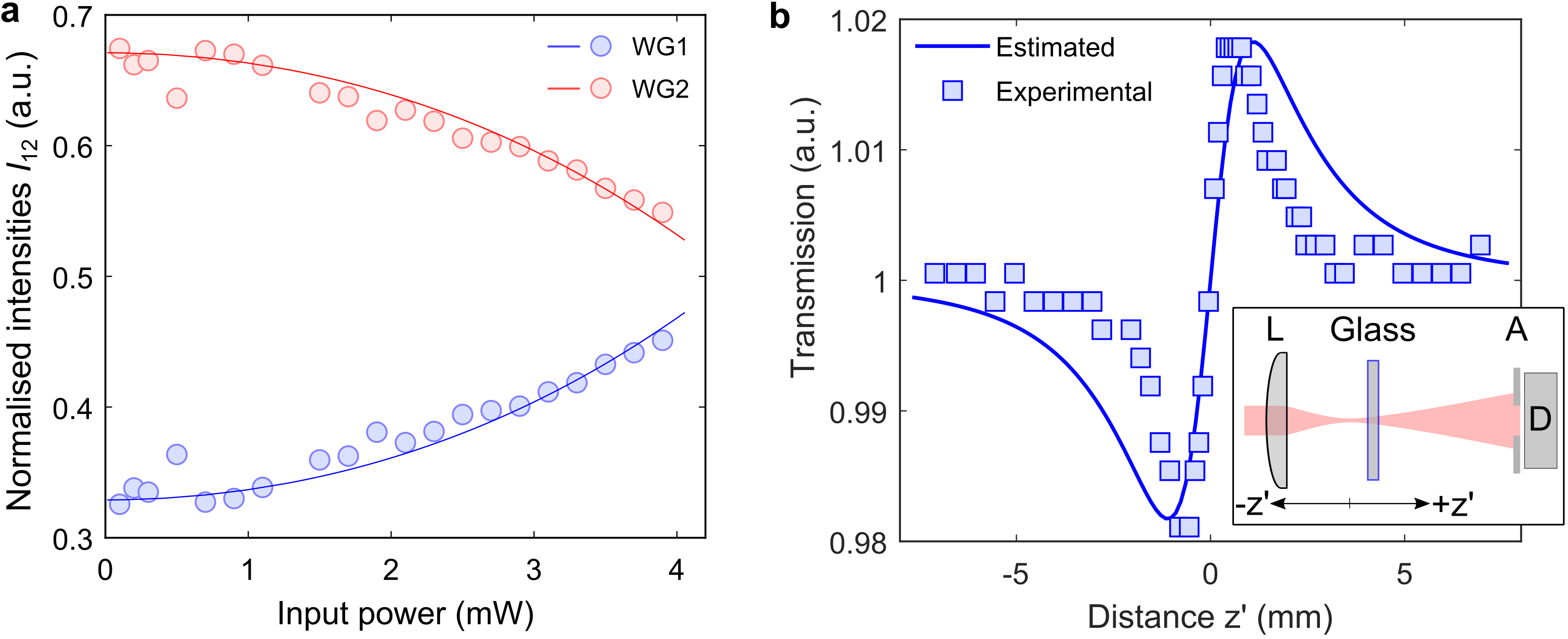}
\caption{{\bf Estimation of renormalised power ${\cal P}$. a,} Characterisation of a nonlinear directional coupler. The data in blue is the measured normalised intensity $I_1$ at the output of waveguide 1 where the light was launched at the input. {\bf b,} Data obtained from the Z-scan of a 1~mm thick borosilicate  sample at 1030~nm wavelength, indicating the self-focusing due to the Kerr effect. The solid line is the estimated Z-scan transmittance curve. The inset shows the experimental setup: here, L is a plano-convex lens, A is an aperture and D is a detector for measuring transmitted optical power.
}
\label{Supp_Fig_n2}
\end{figure*}

The evanescent coupling strength $J$ is a function of wavelength $\lambda$, and as a result it is imperative that the spectrum of the pulse stays within a range with minimal coupling variation. To estimate the variation of coupling, we performed the following experiments. Three sets of two-waveguide couplers were fabricated with $14, 15$ and $16 \, \mu$m inter-waveguide spacing. The evanescent coupling strengths for these couplers were measured as a function of wavelength. For these measurements, broadband light was generated using a commercially available supercontinuum source (NKT Photonics) and a narrow bandwidth ($\!<\!4$~nm) was selected by placing a tunable monochromator after the supercontinuum source. The measured variation of $J(\lambda)$ is shown in Fig.~\ref{Supp_fig2}c. 
For the wavelength range $1030\pm \!10$~nm, $J$ can change only by $\Delta J/J\!<\! \pm4$~\% which is of the order of small fluctuation of $J$ due to the off-diagonal disorder. 
Hence, the new wavelengths generated due to self-phase modulation do not significantly affect the coupling strength. We note that the nonlinear parameter $g$ (defined as $g\!=\!2\pi n_2/(\lambda A_{\text{eff}})$, where $n_2$ is the nonlinear refractive index coefficient and $A_{\text{eff}}$ is the effective area of the waveguide mode) is also a function of wavelength, however, its estimated variation ($\Delta g/g\!=\!\Delta \lambda/\lambda$) is only $\!\sim\!2$\% over $1030\pm \!10$~nm wavelength range.

\begin{figure*}[htb]
\center
\includegraphics[width=12.0 cm]{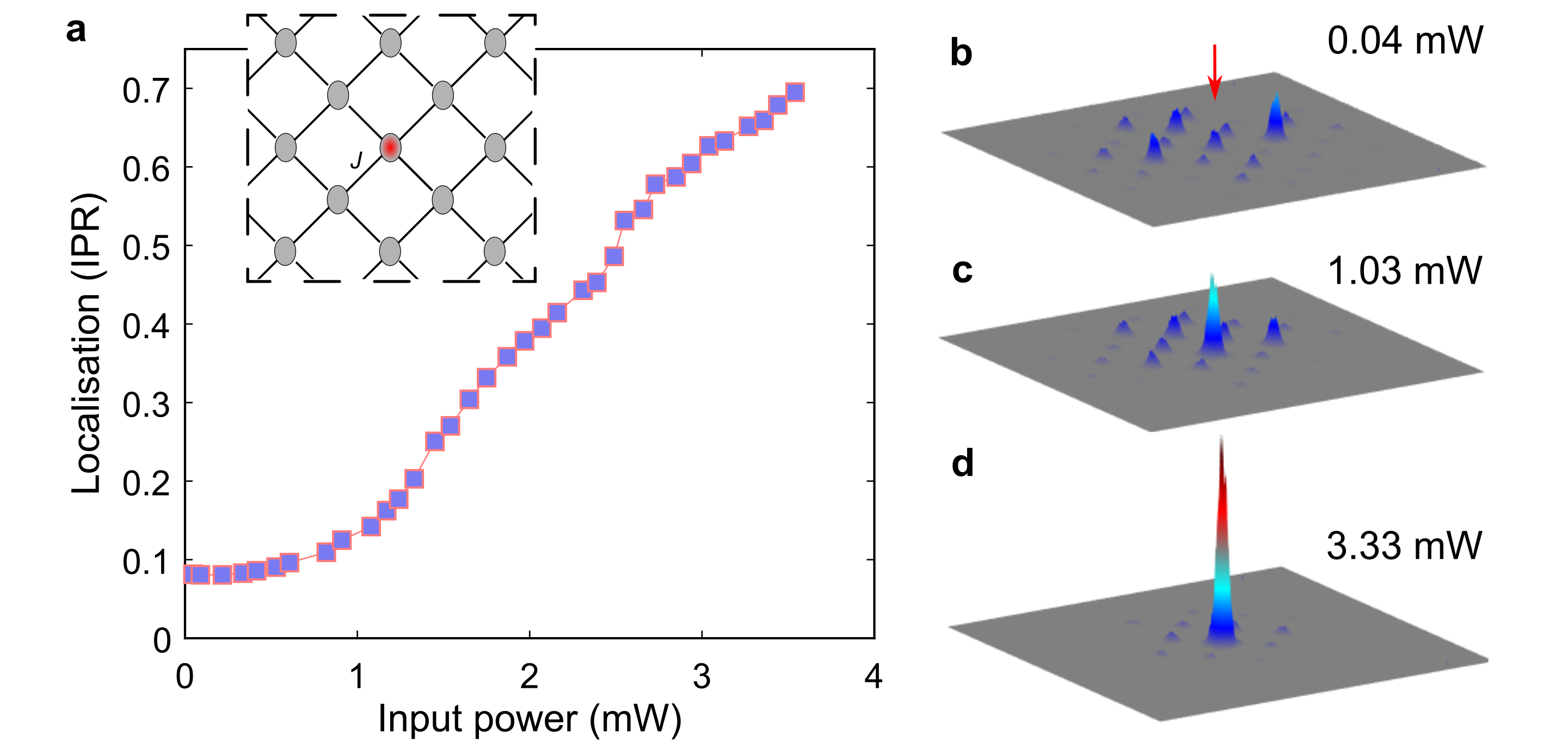}
\caption{{\bf Experimental observation of bulk solitons in a static square lattice. a,} Variation of inverse participation ratio IPR (a measure of localisation) as a function of average input power. The IPR is calculated from the measured intensity distribution at the output of the square lattice. The inset shows the lattice geometry with homogeneous coupling strength $J$. For all measurements, the light was coupled into a single bulk site indicated in red. {\bf b-d,} Output intensity patterns for three different input powers indicated on each image. As a function of input power, the output intensity pattern becomes increasingly localised and finally, all the optical power is trapped in (primarily) one single site [see (d)] where the light was launched at the input (indicated by the red arrow). 
}
\label{Supp_Fig4_staticSqr}
\end{figure*}

\section{3. Estimation of renormalised power ${\cal P}$} \label{nl-power}
The renormalised power ${\cal P}$ was estimated by characterising  a two-waveguide directional coupler and the sign of nonlinearity was determined by the `Z-scan' technique~\cite{sheik1990sensitive} as detailed below.\\

The evolution of light waves in a two-waveguide nonlinear directional coupler is governed by the following equation~~\cite{jensen1982nonlinear}
\begin{eqnarray}
\label{NLcoupler}
i\frac{\partial}{\partial z} \phi_{1,2}(z)\!=\!-J \phi_{2,1} -|\phi_{1,2}|^2 \phi_{1,2}-i\alpha  \phi_{1,2}\; ,
\end{eqnarray}
where $\alpha$ is a measure of (linear) optical losses and $|\phi_{1,2}|^2$ is proportional to the optical power at waveguide 1 or 2, respectively. We aim to estimate ${\cal P}$ at the input i.e., ${\cal P}\!=\!|\phi_{1}|^2+|\phi_{2}|^2$ at $z=0$.
This is done by launching light into one of the waveguides (say WG1) and by measuring the output intensities as a function of input power. 

A directional coupler, consisting of two straight waveguides, was fabricated with $10$~mm length and $17\, \mu$m inter-waveguide separation. The coupling strength of this coupler was measured to be $J\!=\!0.095$~mm$^{-1}$. Fig.~\ref{Supp_Fig_n2}a shows the measured normalised output intensities $I_{1,2}=|\phi_{1,2}|^2/(|\phi_{1}|^2+|\phi_{2}|^2)$ as a function of input power. The solid lines were obtained by solving Eq.~\eqref{NLcoupler} and fitting ${\cal P}$ at the input to be $0.076$~mm$^{-1}$ for unit input power in mW. It should be mentioned that ${\cal P}$ is not conserved during propagation when optical losses are present. 

The above-mentioned characterisation of a nonlinear coupler only provides the absolute value of nonlinearity. To determine whether the glass substrate has a self-focusing (positive) or self-defocusing (negative) nonlinearity, we used `Z-scan' technique~\cite{sheik1990sensitive}. As shown in the inset of Fig.~\ref{Supp_Fig_n2}b, a $1$~mm thick borosilicate glass sample was translated axially (along the $z'$ direction) through the focus of a plano-convex lens with $f\!=\!100$~mm focal length. The transmittance through a finite small aperture (A) in the far-field was measured
as a function of sample position with respect to the beam waist ($z'\!=\!0$).
The beam waist, as well as the intensity of the laser beam, varies along the propagation distance and due to the Kerr effect, the incident radiation can alter the refractive index of a nonlinear medium. In this case, as a function of $z'$, the transmission is expected to show a valley followed by a peak for the positive nonlinearity and a peak followed by a valley for the negative one. As shown in  Fig.~\ref{Supp_Fig_n2}b, a valley followed by a peak was clearly observed in our experiment indicating the positive (self-focusing) nonlinearity.

\section{4. Observation of solitons in a static square lattice} \label{sol-sqr}
In this section, we discuss the observation of bulk solitons in a topologically trivial static square lattice~\cite{szameit2006two}. A $76$-mm-long photonic square lattice with straight optical waveguides was fabricated with coupling strength $J\!=\!0.011$~mm$^{-1}$. In the linear regime, the propagation of light waves across this lattice is governed by a static (i.e.,~$z$-independent) Hamiltonian. For all measurements, the light was coupled into a bulk waveguide away from the edges, hence, the edge effects can be neglected. At low optical power, a significant spread (i.e.,~spatial extent) of the output intensity pattern was observed (see Fig.~\ref{Supp_Fig4_staticSqr}b).
As input power was increased, the output intensity patterns became increasingly localised (Fig.~\ref{Supp_Fig4_staticSqr}a) and finally, all the optical power was  trapped largely in one single site where the light was launched at the input (Fig.~\ref{Supp_Fig4_staticSqr}d). This experiment clearly demonstrates the formation of highly localised solitons in a topologically trivial energy gap of the static square lattice.  This acts as a benchmark showing that we have the ability observe highly-localised solitons in a standard, topologically trivial lattice.

\begin{figure}[htb]
\center
\includegraphics[width=8.6 cm]{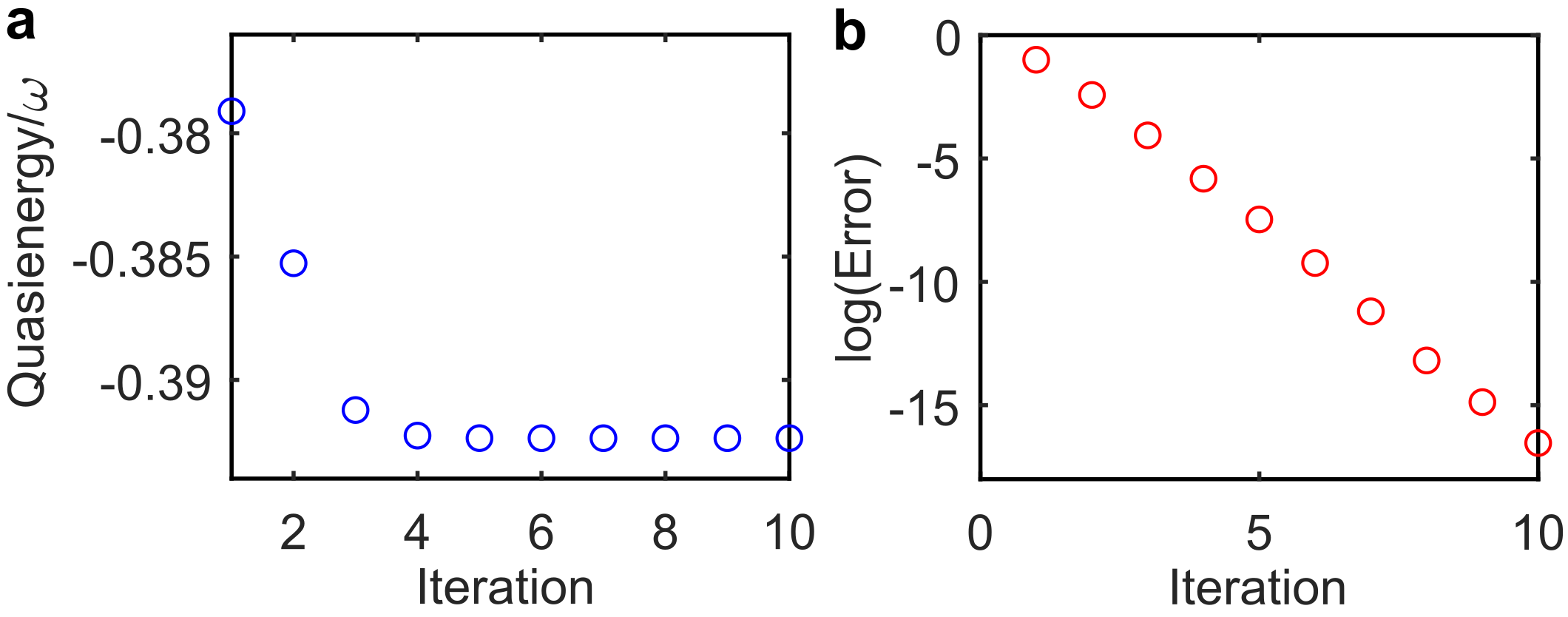}
\caption{{\bf Self-consistency iteration process. a,} Convergence of quasienergy for a soliton at renormalised power ${\cal P}\!=\!0.088$~mm$^{-1}$. Here, $\omega\!=\!2\pi/z_0$. 
{\bf b} Calculated error (defined as $\sum|\phi_s^{n}-\phi_s^{n+1}|^2/{\cal P}$) in each step of iteration.
}
\label{Supp_Fig_sol_itr}
\end{figure}

\section{5. More details on topological solitons}\label{toposol}
As mentioned in the main text, soliton solutions were obtained using the Floquet self-consistency
method~\cite{lumer2013self} where self-localised nonlinear solutions of Eq.~\eqref{nlse} are iteratively calculated starting from an initial guess. 
The linear Hamiltonian of the Floquet topological insulator is periodic in $z$ with periodicity $z_0$. Since we are looking for nonlinear solutions with the same periodicity, the solitons should reproduce themselves after each complete period (up to a phase factor). However, they are allowed to change within each period (this is referred to as `micromotion').

\begin{figure}[htb]
\center
\includegraphics[width=8.6 cm]{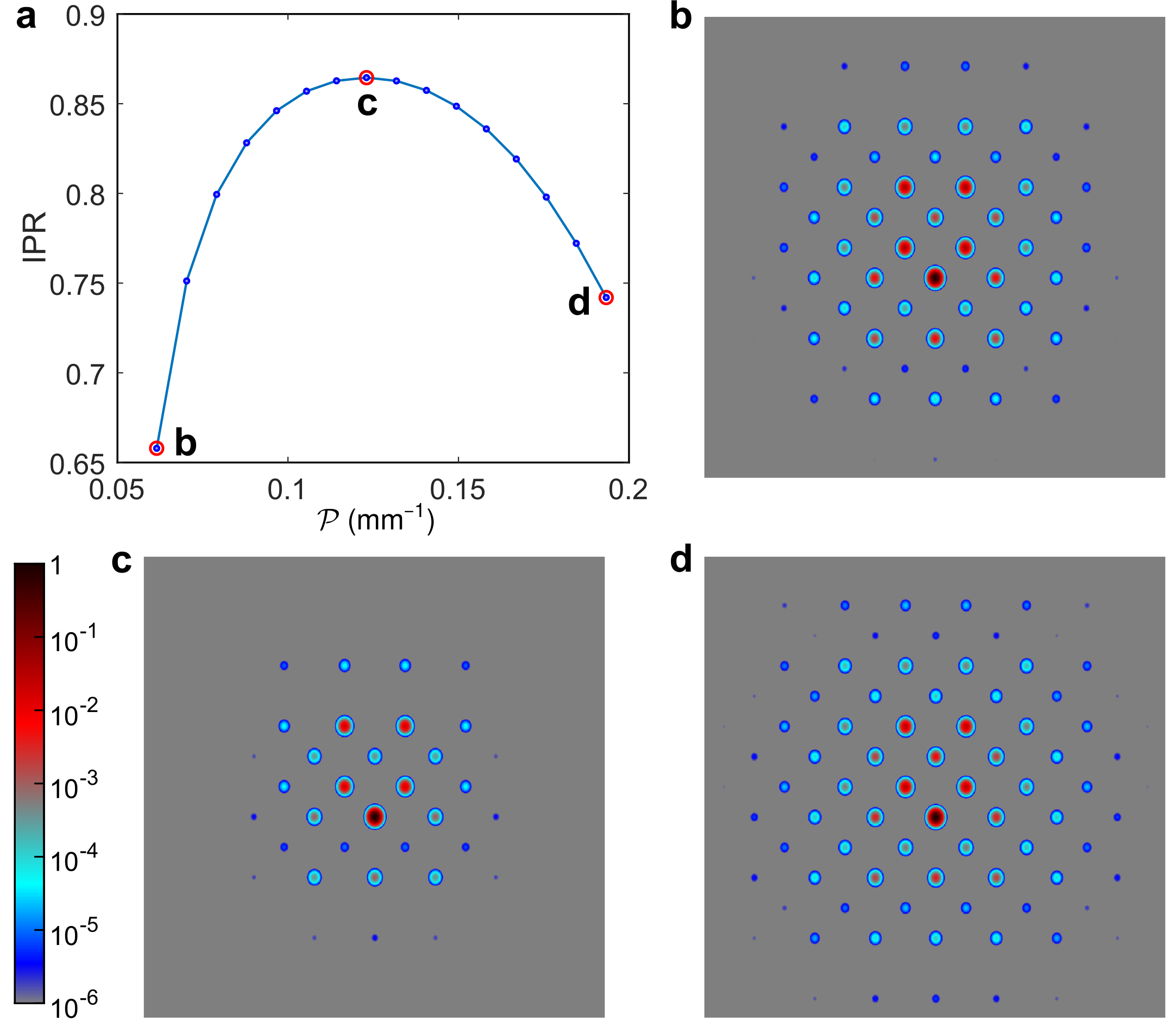}
\caption{{\bf Spatial extent of the band gap solitons. a,} Variation of the spatial extent (size) of the solitons as a function of renormalised power ${\cal P}$.
The soliton solutions were obtained from the self-consistency method. Inverse participation ratio (IPR) is a measure of the localisation of a wavefunction. Here we use IPR to quantify the size of these solitons. {\bf b-d,} Normalised intensity profile of the solitons at three different values of ${\cal P}$ indicated in (a) by red circles. Note that the solitons near the mid-gap quasienergy $\varepsilon\!=\!\pi/z_0$ (i.e.~${\cal P}\!\sim\!0.12$~mm$^{-1}$) have the maximal IPR i.e.,~minimal spatial extent. The field of view is smaller than the actual lattice size.
}
\label{Supp_Fig_sol_size}
\end{figure}

We consider an anomalous Floquet topological insulator (Fig.~\ref{fig1}a) with $200$ lattice sites and periodic boundary conditions. For a chosen value of ${\cal P}\!=\!\sum|\phi_s|^2$, the initial guess solution $\phi_s^{n}(z\!=\!0)$, localised at a single site, is propagated for one complete period of driving to obtain $\phi_s^{n}(z)$ (the superscript ``$n$" indicates iteration step). We then treat $\phi_s^{n}(z)$ as a periodic function of $z$ such that the nonlinear part of the Hamiltonian associated with Eq.~\eqref{nlse} is $z$-periodic. Now one can use Floquet theory to evaluate nonlinear eigenvalues (quasienergy) and eigenfunctions in the presence of a new linear potential defined by the wavefunction $\phi_s^{n}(z)$. We then find the nonlinear eigenfunction $\phi_s^{n+1}$ which has maximum overlap with the normalised initial guess. In the next step of iteration, correctly normalised $\phi_s^{n+1}$ is considered as the initial state. This iteration process is continued until the quasienergy exhibits a convergence  (Fig.~\ref{Supp_Fig_sol_itr}) and the error (defined as $\sum|\phi_s^{n}-\phi_s^{n+1}|^2/{\cal P}$) is less than a threshold which is sufficiently small ($10^{-12}$).

The maximum optical power of these solitons was found to be highly peaked on a single site, irrespective of quasienergy. During propagation, these solitons continuously rotate, executing a cyclotron-like motion, and periodically reproduce themselves.  Fig.~\ref{Supp_Fig_sol_size} shows  the variation of the spatial extent (size) of these solitons as a function of renormalised power ${\cal P}$.  Inverse participation ratio (IPR) is a measure of the localisation of a wavefunction: IPR$=\!1$ means that the wavefunction is localised at a single site.  Normalised intensity profile of the solitons at three different values of ${\cal P}$ is presented in Fig.~\ref{Supp_Fig_sol_size}b-d. Note that the solitons that have quasienergies near mid gap $\varepsilon\!=\!\pi/z_0$ (i.e.~${\cal P}\!\sim\!0.12$~mm$^{-1}$) have the maximal IPR i.e.,~minimum spatial extent, as described in the main text.

\section{6. Description of the supplementary animations~\cite{toposolAnimations}}
Animation1.gif: Propagation of a topological soliton at ${\cal P}\!=\!\pi/z_0$ for one complete period of driving. Note that the soliton continuously rotates performing a cyclotron-like motion and periodically reproduces itself. The colour map is in log scale.\\

Animation2.gif: Variation of the spatial extent (i.e. size) of the topological band gap solitons as a function of renormalised power ${\cal P}$. Note that the solitons near the mid gap quasienergy $\varepsilon\!=\!\pi/z_0$ (i.e.~${\cal P}\!\sim\!0.12$~mm$^{-1}$) have minimal size.\\

Animation3.gif: Experimental observation of bulk solitons in a static square lattice consisting of straight waveguides. (Left) Intensity distribution at the output of a $76$~mm long straight square lattice as a function of (average) input power. The output intensity pattern becomes increasingly localised and finally, all the optical power is trapped largely in one single site where the light was launched at the input.
(Right) Variation of inverse participation ratio IPR (a measure of localisation) as a function of average input power.\\

Animation4.gif:  Experimental observation of topological solitons. 
(Left) Output intensity distributions at $z\!=\!2z_0$ as a function of (average) input power. (Right)  Inverse participation ratio as a function of average input power. This variation of IPR (i.e.~delocalisation to localisation to delocalisation) is qualitatively different than in a topologically trivial static lattice.\\

Animation5.gif:  Experimentally measured output intensity distributions and IPR as a function of input power at $z\!=\!1.5z_0$ of the topological lattice. Otherwise similar to Animation 4.

\newpage
\end{document}